\documentclass[longauth]{aa} 
\usepackage{graphicx}
\usepackage{graphics}
\usepackage{amssymb}
\usepackage{txfonts}
\usepackage{array}
\usepackage{ulem}
\usepackage{longtable}
\usepackage{array} 
\usepackage{tabularx}
\usepackage{natbib}
\bibpunct{(}{)}{;}{a}{}{,}   
\begin{document}
\title{VVV  DR1: The First  Data Release  of the  Milky Way  Bulge and
  Southern  Plane  from  the  Near-Infrared ESO  Public  Survey  VISTA
  Variables  in the  V\'{i}a  L\'actea\thanks{Based on  observations
    taken  within  the  ESO  VISTA  Public Survey  VVV,  Programme  ID
    179.B-2002}}
%
{   
  \author{R.~K.~Saito$^{1}$
   \and M.~Hempel$^{1}$
   \and D.~Minniti$^{1,2,3}$
   \and P.~W.~Lucas$^{4}$
   \and M.~Rejkuba$^{5}$ 
   \and I.~Toledo$^{6}$
   \and O.~A.~Gonzalez$^{5}$  
   \and J.~Alonso-Garc\'{i}a$^{1}$  
   \and M.~J.~Irwin$^{7}$
   \and E.~Gonzalez-Solares$^{7}$
% ...
}
}

\offprints{R. K. Saito: rsaito@astro.puc.cl} 

\institute{ 
$^{1}$Departamento   Astronom\'ia   y  Astrof\'isica,   Pontificia Universidad  Cat\'olica  de Chile,  Av.  Vicu\~na Mackenna  4860, Santiago,   Chile\\ 
$^{2}$Vatican Observatory, Vatican City State V-00120, Italy\\
$^{3}$Department of Astrophysical Sciences, Princeton University, Princeton, NJ 08544-1001, USA\\
$^{4}$Centre for Astrophysics Research, University of Hertfordshire, College Lane, Hatfield AL10 9AB, UK\\
$^{5}$European Southern  Observatory, Karl-Schwarzschild-Strasse 2, D-85748 Garching, Germany\\
$^{6}$Atacama Large Millimeter Array, Alonso de C\'ordova 3107, Vitacura, Santiago, Chile\\
$^{7}$Institute  of Astronomy,  University of  Cambridge, Madingley Road, Cambridge CB3 0HA, UK\\
$^{8}$Institute for Astronomy, The University of Edinburgh, Royal Observatory, Blackford Hill, Edinburgh EH9 3HJ, UK\\
$^{9}$European Southern  Observatory, Ave. Alonso  de Cordova 3107, Casilla 19, Santiago 19001, Chile\\
$^{10}$Jodrell Bank Centre for Astrophysics, The University of Manchester, Oxford Road, Manchester M13 9PL, UK\\
$^{11}$Astronomy Unit, School of Physics and Astronomy, Queen Mary University of London, Mile End Road, London, E1 4NS, UK\\
$^{12}$Departamento de F\'{i}sica, Universidad de La Serena, Cisternas 1200 Norte, La Serena, Chile\\
$^{13}$Faculdade de Ci\^encias da Universidade de Lisboa, Campo Grande, Edificio C5, 1749-016 Lisboa, Portugal\\
$^{14}$Laborat\'orio Nacional de Astrof\'{i}sica, Rua Estados Unidos 154, Itajub\'a-MG,  37504-364, Brazil\\
$^{15}$Instituto de Astronom\'{i}a Te\'orica y Experimental, CONICET, Laprida 922, 5000 C\'ordoba, Argentina\\
$^{16}$Centro de Astro-Ingenier\'{i}a, Pontificia Universidad  Cat\'olica  de Chile,  Av.  Vicu\~na Mackenna  4860, Santiago,   Chile\\ 
$^{17}$INAF - Osservatorio Astronomico di Bologna, via Ranzani 1, 40127, Bologna\\
$^{18}$Warsaw University Observatory, Al. Ujazdowskie 4,00-478, Warsaw, Poland\\
$^{19}$Departmento de Astronom\'ia, Universidad de Concepci\'on, Casilla 160-C, Concepci\'on, Chile\\
$^{20}$Departamento de F\'{i}sica y Astronom\'{i}a, Facultad de Ciencias, Universidad de Valpara\'{i}so, Ave. Gran Breta\~na 1111, Playa Ancha, Casilla 5030, Valpara\'{i}so, Chile\\
$^{21}$Observatorio Astron\'omico de C\'ordoba, Universidad Nacional de C\'ordoba, Laprida 854, 5000 C\'ordoba, Argentina\\
$^{22}$Consejo Nacional de Investigaciones Cient\'{i}ficas y T\'ecnicas, Av. Rivadavia 1917 - CPC1033AAJ - Buenos Aires, Argentina\\
$^{23}$Gemini Observatory, Southern Operations Center, c/o AURA, Casilla 603 La Serena, Chile\\
$^{24}$Department of Astronomy, University of Florida, 211 Bryant Space Science Center P.O. Box 112055, Gainesville, FL, 32611-2055, USA\\
$^{25}$Instituto de Ciencias Astron\'omicas, del la Tierra y del Espacio (ICATE-CONICET), Av. Espa\~na Sur 1512, J5402DSP San Juan, Argentina\\
$^{26}$Universidade de S\~ao Paulo, IAG, Rua do Mat\~ao 1226, Cidade Universit\'aria, S\~ao Paulo 05508-900, Brazil\\
$^{27}$Facultad de Ciencias Astron\'omicas y Geof\'{i}sicas, Universidad Nacional de La Plata, and Instituto de Astrof\'isica La Plata (IALP--CONICET), Paseo del Bosque S/N, B1900FWA, La Plata, Argentina\\
$^{28}$INAF - Astronomical Observatory of Padova, vicolo dell Osservatorio 5, I - 35122 Padova, Italy\\ 
$^{29}$Dipartimento di Astronomia, Universit\'{a} di Padova, vicolo dell Osservatorio 3, 35122 Padova, Italy\\
$^{30}$Department of Physics, University of Wisconsin-Whitewater, 800 West Main Street, Whitewater, WI 53190\\
$^{31}$Universidade Federal do Rio Grande do Sul, IF, CP 15051, Porto Alegre 91501-970, RS, Brazil\\
$^{32}$Departamento de Astronom\'{i}a, Universidad de Chile, Casilla 36-D, Santiago, Chile\\
$^{33}$Kavli Institute for Astronomy and Astrophysics, Peking University, Yi He Yuan Lu 5, Hai Dian District, Beijing 100871, China\\
$^{34}$Department of Astronomy and Space Science, Kyung Hee University, Yongin-shi, Kyungki-do 449-701, Republic of Korea\\
$^{35}$Max Planck Institute for Astronomy, K\"onigstuhl 17, 69117 Heidelberg, Germany\\
$^{36}$Departamento de F\'{\i}sica, Ingenier\'{\i}a de Sistemas y Teor\'{\i}a de la Se\~{n}al, Universidad de Alicante, Apdo. 99, E03080 Alicante, Spain\\
$^{37}$Mullard Space Science Laboratory, University College London, Holmbury St. Mary, Dorking, Surrey, RH5 6NT, UK\\
$^{38}$Department of Physics, University of Cincinnati, PO Box 210011, Cincinnati, OH 45221-0011, USA\\
$^{39}$School of Physics \& Astronomy, University of Leeds, Woodhouse Lane, Leeds LS2 9JT, UK\\
$^{40}$NASA-Ames Research Center/Bay Area Environmental Research Institute, MS 244-30, Moffett Field, CA 94035, USA\\
$^{41}$UK Astronomy Technology Centre, Royal Observatory, Blackford Hill, Edinburgh EH9 3HJ\\
$^{42}$Instituto de Astrof\'{i}sica de Canarias, V\'{i}a L\'actea s/n, E38205 - La Laguna (Tenerife), Spain\\
$^{43}$Departamento de Astrof\'{i}sica, Universidad de La Laguna, E-38206, La Laguna, Tenerife, Spain\\
$^{44}$School of Physics and Astronomy, University of Southampton, Highfield, Southampton, SO17 1BJ, UK\\
$^{45}$Saint Marys University, 923 Robie Street, Halifax, Nova Scotia, Canada\\
$^{46}$Centro de Astrobiologia CSIC - INTA, Carretera Torrej\'on-Ajalvir km 4, 28850, Madrid, Spain\\
$^{47}$Istituto di Astrofisica Spaziale e Fisica Cosmica di Bologna, via Gobetti 101, 40129 Bologna, Italy\\
$^{48}$Service d'Astrophysique - IRFU, CEA-Saclay, 91191 Gif sur Yvette, France\\
$^{49}$Instituto de Astronom\'{i}a y F\'{i}sica del Espacio, Casilla de Correo 67, Sucursal 28, Buenos Aires, Argentina\\
$^{50}$Department of Physics \& Astronomy, Macquarie University, Sydney, NSW 2109, Australia\\
$^{51}$Australian Astronomical Observatory, PO Box 296, Epping, NSW 1710, Australia\\
$^{52}$Departamento de Ciencias Fisicas, Universidad Andres Bello, Av. Republica 252, Santiago, Chile\\
$^{53}$South African Astronomical Observatory, PO Box 9, Observatory 7935, Cape Town, South Africa\\
$^{54}$Hartebeesthoek Radio Astronomy Observatory, PO Box 443, Krugersdorp 1740, South Africa\\
$^{55}$Aryabhatta Research Institute of Observational Sciences (ARIES), Manora Peak, 263 129 Nainital, India\\
$^{56}$The University of Kent, Canterbury, Kent, CT2 7NH, UK\\
$^{57}$Gemini Observatory, Northern Operations Center, 670 N. A'ohoku Place, Hilo, Hawaii, 96720, USA\\
$^{58}$Division of Optical and Infrared Astronomy, National Astronomical Observatory of Japan 2-21-1 Osawa, Mitaka, Tokyo, 181-8588, Japan\\
$^{59}$Departamento de Ingenier\'{i}a El\'ectrica, Pontificia Universidad Cat\'olica de Chile, Av. Vicu\~na Mackenna 4860, Santiago, Chile\\
}
  
   \date{Received ; Accepted }

  
\abstract
{The ESO Public  Survey VISTA Variables in the  V\'{i}a L\'actea (VVV)
  started in 2010.  VVV targets 562~sq.~deg in the  Galactic bulge and
  an adjacent plane region and is expected to run for $\sim 5$ years.}
{In this paper we describe  the progress of the survey observations in
  the first  observing season, the  observing strategy and  quality of
  the data obtained.  }
{The observations  are carried out on  the 4-m VISTA  telescope in the
  $ZYJHK_{\rm s}$ filters.  In  addition to the multi-band imaging the
  variability  monitoring  campaign  in  the $K_{\rm  s}$  filter  has
  started.  Data  reduction is carried  out using the pipeline  at the
  Cambridge Astronomical Survey  Unit. The photometric and astrometric
  calibration is performed via  the numerous 2MASS sources observed in
  each pointing.}
{The  first   data  release  contains  the   aperture  photometry  and
  astrometric   catalogues  for  348   individual  pointings   in  the
  $ZYJHK_{\rm  s}$ filters taken  in the  2010 observing  season.  The
  typical  image quality  is  $\sim0\farcs9-1\farcs0$.  The  stringent
  photometric  and  image  quality  requirements  of  the  survey  are
  satisfied in 100\% of the $JHK_{\rm  s}$ images in the disk area and
  90\%  of  the   $JHK_{\rm  s}$  images  in  the   bulge  area.   The
  completeness in the $Z$ and $Y$ images is 84\% in the disk, and 40\%
  in the bulge.  The  first season catalogues contain $1.28\times10^8$
  stellar sources in  the bulge and $1.68\times10^8$ in  the disk area
  detected in  at least one  of the photometric bands.   The combined,
  multi-band  catalogues contain  more  than $1.63\times10^8$  stellar
  sources.   About  10\%  of   these  are  double  detections  due  to
  overlapping   adjacent   pointings.    These  overlapping   multiple
  detections are  used to characterise  the quality of the  data.  The
  images  in the  $JHK_{\rm s}$  bands extend  typically $\sim  4$ mag
  deeper  than 2MASS.   The  magnitude limit  and photometric  quality
  depend  strongly on  crowding in  the inner  Galactic  regions.  The
  astrometry for $K_s=15-18$~mag has $rms \sim 35-175$~mas.}
{The  VVV Survey  data  products offer  a  unique dataset  to map  the
  stellar populations in the Galactic bulge and the adjacent plane and
  provide an exciting new tool for the study of the structure, content
  and  star  formation   history  of  our  Galaxy,  as   well  as  for
  investigations of  the newly discovered star  clusters, star forming
  regions in the disk,  high proper motion stars, asteroids, planetary
  nebulae, and other interesting objects.}

\authorrunning{Saito   et   al.}
\titlerunning{VVV DR1} 
\maketitle
%
\section{Introduction}

The VISTA  Variables in the  V\'{i}a L\'actea (VVV) Survey  is mapping
562 square degrees in the Galactic  bulge and the southern disk in the
near-infrared  \citep{2010NewA...15..433M}.    The  VVV  Survey  gives
near-IR multi-colour information in five passbands: $Z$ (0.87~$\mu$m),
$Y$ (1.02~$\mu$m),  $J$ (1.25~$\mu$m), $H$  (1.64~$\mu$m), and $K_{\rm
  s}$ (2.14~$\mu$m), as  well as time coverage spanning  over 5 years,
that will complement past/recent, current and upcoming surveys such as
2MASS, DENIS, GLIMPSE-II, VPHAS+,  MACHO, OGLE, EROS, MOA, PLANET, and
GAIA.

VVV is an ESO Public Survey, i.e., the observational raw data are made
available  to  the  astronomical  community immediately,  whereas  the
reduced data will  be published once a year in  a Data Release through
ESO.  This  paper describes and  characterises the first  data release
(DR1) of the VVV Survey. Some first results of the VVV Survey based on
early  science  images  taken  in   the  bulge  and  disk  fields  are
highlighted          in         \cite{2010Msngr.141...24S}         and
\cite{2011rrls.conf..145C}.

The preparatory  phase for  the VVV Survey  started in 2006,  with the
first test  observations obtained in October  2009. Regular operations
started with the first survey observations in February 2010.  The data
collected during the first year of observations until October 2010 are
the  subject of  this public  release.  Our  survey is  planned  to be
carried out for  5 years, and we expect  to produce yearly accumulated
data releases.

The VVV Survey is foremost a variability study of the inner regions of
the  Milky Way \citep{2010NewA...15..433M},  but will  also complement
the  existing   2MASS  $JHK$  photometry  \citep{2003tmc..book.....C},
extending to  much fainter limits while adding  two additional filters
($ZY$), and  providing time domain information  useful for variability
and proper motion studies. In particular, the higher resolution of the
VVV data represents a huge  advantage in crowded fields in relation to
previous    near-IR    surveys     such    as    2MASS    and    DENIS
\citep{1994Ap&SS.217....3E},  where  the  single-epoch photometry  was
confusion-limited,  reaching  $K_{\rm  s}\sim14.3$~mag.  The  limiting
magnitude  of  the  VVV  data  using aperture  photometry  is  $K_{\rm
  s}\sim18.0$~mag  in  most  fields.  Even  in  the  innermost  fields
($|b|\leq 1^\circ$) the VVV Survey reaches $K_{\rm s}\sim16.5$~mag, at
least a magnitude  deeper than the IRSF/SIRIUS Survey  of the Galactic
Centre
\citep{2003SPIE.4841..459N,2006ApJ...647.1093N,2009ApJ...696.1407N}.

The   VVV   Survey  was   designed   to   complement  the   UKIDSS-GPS
\citep{2008MNRAS.391..136L},                                     VPHAS+
\citep[see][]{2007Msngr.127...28A},   and   the   GLIMPSE-II   surveys
\citep{2003PASP..115..953B}.   The  UKIDSS-GPS  is  mapping  $|{b}|  <
5^\circ$  in Galactic  latitude in  the northern  plane for  3 epochs,
while  VPHAS+ also  observes the  Galactic  plane in  the optical  and
H$\alpha$ using the ESO VLT Survey Telescope (VST).

GLIMPSE-II survey images  the central $\pm10^\circ$ of the  plane in 4
bands with IRAC. VVV  provides variability information for the overlap
region, supporting  studies of the content and  distribution of stars,
stellar populations, and interactions  of the strong nuclear wind with
the ambient ISM  above and below the nucleus, as well  as the rate and
location of current star formation.

Multiband  {\it   Spitzer}  public   surveys  with  IRAC   (mid-IR  at
3.6~$\mu$m, 4.5~$\mu$m,  5.6~$\mu$m and 8.0~$\mu$m)  and MIPS (far-IR,
23.7~$\mu$m,  71.4~$\mu$m weighted average  wavelength), respectively,
cover the mid-plane  $(|b|<1^\circ)$ at $65^\circ < l  < 10^\circ$ and
$-10^\circ  < l  < -65^\circ$.   The  southern half  of these  surveys
overlaps  with   the  VVV  disk  area,  allowing   the  detection  and
characterisation of star formation  regions and to probe the structure
of the  inner disk of the  Galaxy.  In optically  obscured regions the
IRAC data complement both the VVV Survey and the VST/VPHAS+ by tracing
the influence of the most massive stars on star formation.

In  addition, we  complement the  existing bulge  microlensing surveys
such     as     MACHO     \citep{2000ApJ...541..734A}     and     OGLE
\citep{1993AcA....43...69U},  which  observe  in optical  bands,  with
limited or no colour  information. These surveys mostly concentrate on
regions  of  low  extinction.   VVV will  provide  useful  variability
information for the overlap regions.

The  data  are of  excellent  quality in  general,  and  only a  small
fraction did not  pass our quality controls and  had to be reacquired.
These  DR1  data have  passed  all  the  initial quality  controls  as
performed  by the  survey  team in  collaboration  with the  Cambridge
Astronomical          Survey         Unit         (CASU)\footnote{{\tt
    http://casu.ast.cam.ac.uk/vistasp/}}.   We checked  image defects,
telescope problems,  seeing, zero point,  magnitude limit, ellipticity
and airmass,  for instance.  However,  it is important to  stress that
the data  quality and calibrations  will improve with  subsequent data
releases.

Here we will address the  general information for the community, e.g.,
on  the  survey area  and  strategy,  data  quality, progress  in  the
observations, published source lists,  as well as examples of specific
applications.   In  addition,  the  data,  procedures  and  additional
information  are  available   through  the  ESO  Archive\footnote{{\tt
    http://www.eso.org/sci/archive.html}}, the VVV Survey Science Team
homepage\footnote{{\tt  http://vvvsurvey.org}}, CASU, and  through the
VISTA      Science       Archive      (VSA)      webpage\footnote{{\tt
    http://horus.roe.ac.uk/vsa/index.html}}.

This paper is organized as follows:

Section~\ref{sec:area_obs} describes the area coverage and the
observations with VISTA as well as the data processing.
Section~\ref{sec:photometry} describes the photometric quality
(limits, accuracy), as well as completeness.
Section~\ref{sec:2mass} discusses a comparison with 2MASS.
Section~\ref{sec:psf} presents a comparison between the VVV DR1
(aperture) catalogues and PSF photometry.
Section~\ref{sec:astrometry} describes the astrometric data quality.
Section~\ref{sec:maps} presents density maps for the bulge and disk
fields.  
Section~\ref{sec:dia} describes a suitability test of the DR1 data for
difference image analysis.

The final  section summarizes and presents  our conclusions, including
important caveats  regarding this DR1,  and future improvements  to be
implemented in  DR2. Finally, the  VVV tile coordinates are  listed in
the Appendix.

\section{Survey Area and Observations}
\label{sec:area_obs}

\subsection{Telescope and Instrument}
\label{sec:telinst}

The telescope used  to carry out the VVV Survey  is VISTA (Visible and
Infrared  Survey  Telescope for  Astronomy),  a  4-m class  wide-field
telescope  with  a  single  instrument, VIRCAM  \citep[VISTA  InfraRed
  CAMera;][]{2006SPIE.6269E..30D,2010SPIE.7733E...4E},  located on its
own peak at ESO's Cerro Paranal Observatory in Chile, about 1500m away
from the VLT.  Its primary mirror  has a diameter of 4.1m, providing a
$f$\,/3.25 focal ratio at the Cassegrain focus where the instrument is
mounted.  The secondary  mirror has  a  1.24m diameter.  The start  of
survey  operations of the  telescope was  on April  1, 2010,  but most
VISTA surveys, including  VVV, started collecting science observations
a  few  months  earlier, in  parallel  with  the  last phases  of  the
scientific   performance  verification   and   operations  fine-tuning
performed  by Paranal  Observatory staff.   During the  first  year of
operations the mirrors were coated with silver, which is optimized for
near-IR observations.

With 1.64 deg diameter VIRCAM  offers the largest unvignetted field of
view in  the near-IR regime on  4-m class telescopes.   It is equipped
with 16  Raytheon VIRGO $2048  \times 2048$ pixels$^2$  HgCdTe science
detectors, with  $0\farcs 339$  average pixel scale.   Each individual
detector therefore covers $\sim 694 \times 694$~arcsec$^2$ on the sky.
The  achieved image quality  (including seeing)  is better  than $\sim
0\farcs6$ on axis. The image quality distortions are up to about 10\%
across the  wide field of  view.  The detectors  are arranged in  a $4
\times  4$  array, with  large  spacings of  90\%  and  42.5\% of  the
detector  size along  the $X$  and  $Y$ axes,  respectively. A  single
pointing, called a pawprint, covers 0.59~sq.~deg, and provides partial
coverage of the field of view.  By combining 6 pawprint exposures with
appropriate offsets a contiguous coverage  of a field is achieved with
at least  2 exposures  per pixel  except at two  edges.  In  all VISTA
observations such a  field is called a tile  and covers a 1.64~sq.~deg
field  of view.   Throughout this  paper  we will  use a  tile as  the
individual exposure.

VIRCAM has four  additional optical CCDs, two for  guiding and two for
active  optics.  For exposures  longer than  $\sim 40$~sec  the active
optics is  run in parallel mode  with the observations.   Due to VVV's
very short  individual exposures of  (see Sec.~\ref{sec:strategy}) the
active  optics correction  is only  performed every  $\sim  30$~min or
after a  larger offset.   This, combined with  the need to  survey the
area  fast (hence minimizing  the overheads  for more  frequent active
optics corrections),  and typical seeing  on Paranal limits  the image
quality   obtained   for   the   survey  data   to   typically   $\sim
0.9-1.0$~arcsec (see Sec.~\ref{sec:reduction}).

\begin{figure}
\centering
\resizebox{\hsize}{!}{
\includegraphics[angle=270]{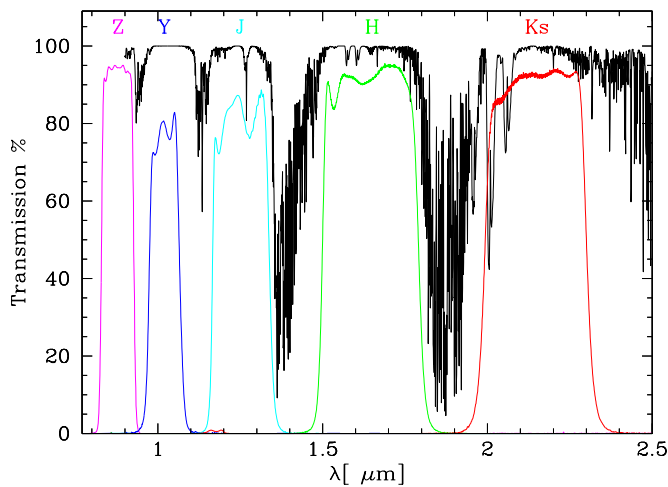}
}
\caption{Transmission curves  for the five broad  band filters present
  at the VIRCAM: $Z$, $Y$, $J$, $H$ and $K_{\rm s}$, compared with the
  typical  atmospheric  transmission profile  for  airmass~$=1.0$  and
  $1.0$~mm water  vapour.  The  effective wavelengths for  all filters
  are listed in Table~\ref{tab:lambda}.}
\label{fig:filters}
\end{figure}

\begin{figure*}[ht]
\includegraphics[bb=1.7cm 13cm 5cm 19cm,scale=1.20]{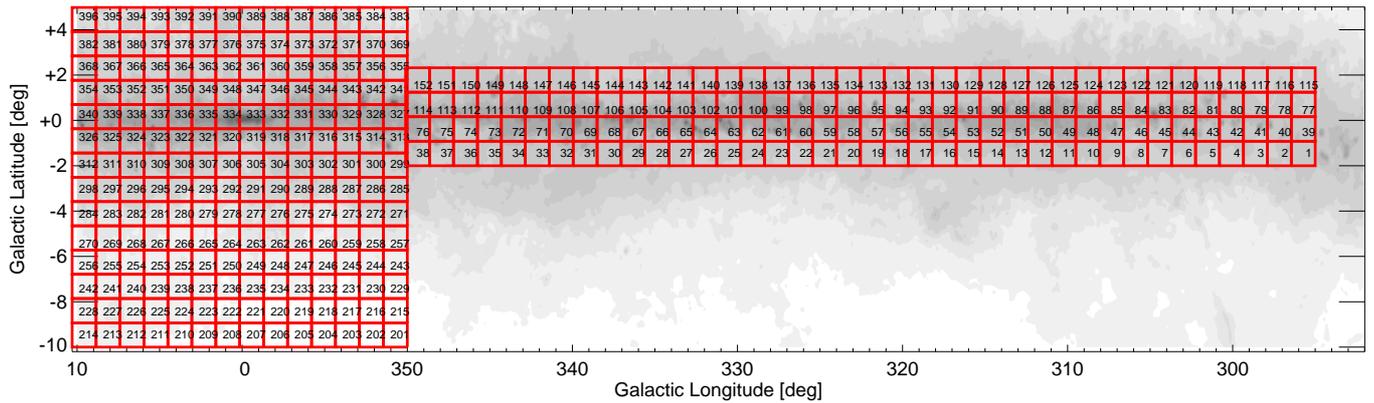}
\caption{VVV Survey area and tile numbering. The tile names start with
  ``b'' for bulge and ``d''  for disk tiles, followed by the numbering
  as  shown in the  figure. The centre  coordinates for  all VVV
    tiles  are  listed  in  Table~\ref{tab:tiles}.}
\label{fig:survey_area}
\end{figure*}

VIRCAM is equipped with 5  broad-band filters ($Z$, $Y$, $J$, $H$, and
$K_{\rm  s}$)  and  two   narrow-band  filters  centred  at  0.98  and
1.18~$\mu$m.  The  VVV Survey uses  all 5 broad band  filters spanning
from  0.84 to  2.5~$\mu$m.  Their  effective wavelengths  and relative
extinctions   are   given   in   Table~\ref{tab:lambda},   while   the
transmission curves are shown in Fig.~\ref{fig:filters}, compared to a
typical atmospheric transmission profile for airmass 1.0 and 1mm water
vapour in the atmosphere.

For more  details about the telescope  and instrument we  refer to the
VIRCAM           instrument           web          pages\footnote{{\tt
    http://www.eso.org/sci/facilities/paranal   /instruments/vircam/}},
and the VISTA/VIRCAM User Manual \citep{ivanov+szeifert09}.

\begin{table}
\caption[]{Effective wavelengths for the VISTA filter set used in the
  VVV observations  and the relative extinction for  each filter based
  on      the      \cite{1989ApJ...345..245C}      extinction      law
  \citep[from][]{2011rrls.conf..145C}.}
\begin{center}
\label{tab:lambda}
\begin{tabular}{cccc}
\hline \hline 
\noalign{\smallskip}
Filter & ${\rm \lambda}_{\rm eff} (\mu m)$ & $A_X/A_V$  & $A_X/E(B-V)$ \\
\noalign{\smallskip}
\hline
\noalign{\smallskip}
$ Z $ & 0.878 & 0.499 & 1.542 \\
$ Y $ & 1.021 & 0.390 & 1.206 \\
$ J $ & 1.254 & 0.280 & 0.866 \\
$ H $ & 1.646 & 0.184 & 0.567 \\
$ K_{\rm s}$ & 2.149 & 0.118 & 0.364 \\
\noalign{\smallskip}
\hline
\end{tabular}
\end{center}
\end{table}

\subsection{Survey area}
\label{sec:surveyarea}

The VVV Survey area consists of  348 tiles, 196 tiles in the bulge and
152  in the disk  area.  These  two components  were planned  to cover
520~sq.~deg,  as  follows:  (i)  the  VVV  bulge  survey  area  covers
300~sq.~deg between  $-10^\circ \leq l \leq  +10^\circ$ and $-10^\circ
\leq  b \leq  +5^\circ$;  and (ii)  the  VVV disk  survey area  covers
220~sq.~deg between  $295^\circ \leq  l \leq 350^\circ$  and $-2^\circ
\leq b \leq  +2^\circ$.  However, in order to  maximize the efficiency
of the  tilling process  (see Section \ref{sec:strategy}),  the Survey
Area Definition Tool \citep[SADT;][]{hilker11} produced some shifts at
the  edges  of  the  survey  area,  and  as  the  result  an  area  of
$\sim$562~sq.~deg  (42~sq.~deg   larger)  was  observed.    Thus,  the
observed area is within  $-10.0^\circ \lesssim l \lesssim +10.4^\circ$
and within $-10.3^\circ \lesssim  b \lesssim +5.1^\circ$ in the bulge,
and  $294.7^\circ \lesssim  l \lesssim  350.0^\circ$  and $-2.25^\circ
\lesssim b \lesssim +2.25^\circ$ in  the disk. The VVV Survey area and
tile numbering are shown in Fig.~\ref{fig:survey_area}, while the list
of all tile centres in Equatorial and Galactic coordinates is given in
Table~\ref{tab:tiles}.  The tile names  start with ``b'' for bulge and
``d''   for  disk   tiles,  followed   by  the   numbering   shown  in
Fig.~\ref{fig:survey_area}.

\begin{table}
\caption[]{VVV Survey completion in the 2010 season.}
\begin{center}
\label{tab:completion}
\begin{tabular}{cccc}
\hline \hline 
\noalign{\smallskip}
Tile & Completed & Total  & Completion \\
type &   Tile    & Tiles  & percentage \\
\noalign{\smallskip}
\hline \\
\noalign{\smallskip}
\multicolumn{4}{c}{Bulge} \\
\hline
\noalign{\smallskip}
$JHK_{\rm s}$ & 188  & 196   &  95\% \\
$ZY$        & 78    & 196   &  40\%  \\
Variability & 113  & $5 \times 196$ & 12\% \\
\noalign{\smallskip}
\hline \\
\multicolumn{4}{c}{Disk} \\
\hline
\noalign{\smallskip}
$JHK_{\rm s}$ & 152 & 152   &  100\% \\
$ZY$        & 128  & 152  &   84\%  \\
Variability & 547  & $5 \times 152$ & 72\% \\
\noalign{\smallskip}
\hline
\end{tabular}
\end{center}
\end{table}

\begin{figure*}[ht]
\includegraphics[bb=.5cm -4.cm 19cm 10cm,angle=-90,scale=0.5]{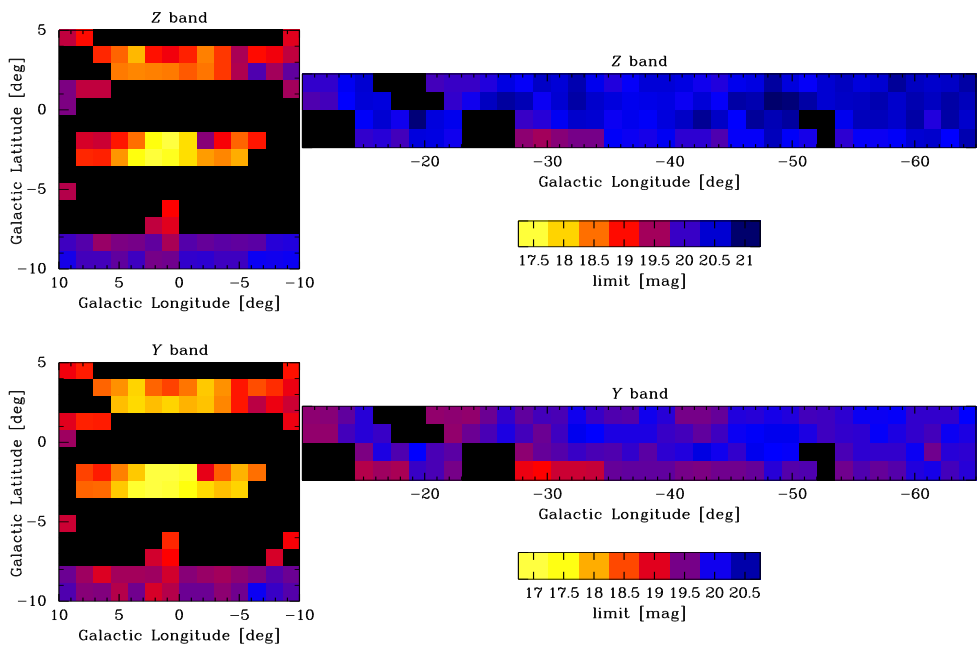}
\caption{The $5\sigma$  magnitude limits of the catalogues  in the $Z$
  (top panel) and $Y$ bands (bottom panel).  The colour scale is shown
  in each  case.  The completeness of  the DR1 can be  also checked in
  the maps, where the missing tiles appear in black. Exposure times in
  the     bulge    and    disk     fields    are     different    (see
  Table~\ref{tab:strategy}), which contributes the disk fields to have
  deeper  photometry.   Similar maps  for  $J$,  $H$  and $K_{\rm  s}$
  photometry, as  well as for the variability  campaign, are presented
  in Figs.~\ref{fig:maglim_jh} and \ref{fig:maglim_kv}.}
\label{fig:maglim_zy}
\end{figure*}

\begin{figure*}[ht]
\includegraphics[bb=.5cm -4.cm 19cm 10cm,angle=-90,scale=0.5]{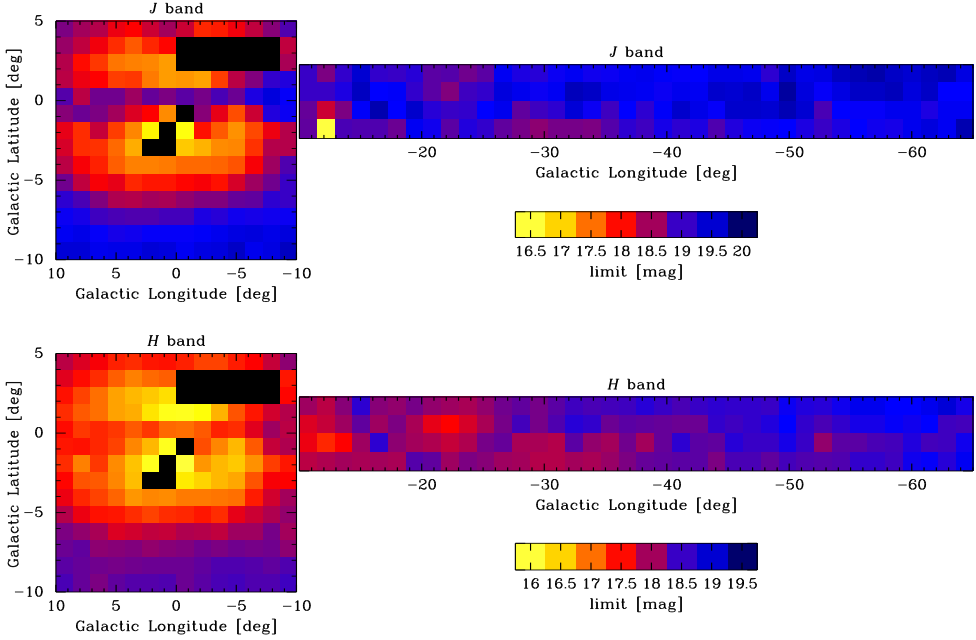}
\caption{The $5\sigma$  magnitude limits of the catalogues  in the $J$
  (top panel) and $H$ bands (bottom panel). The notation is similar to
  that presented in Fig.~\ref{fig:maglim_zy}.}
\label{fig:maglim_jh}
\end{figure*}

\begin{figure*}[ht]
\includegraphics[bb=.5cm -4.cm 19cm 10cm,angle=-90,scale=0.5]{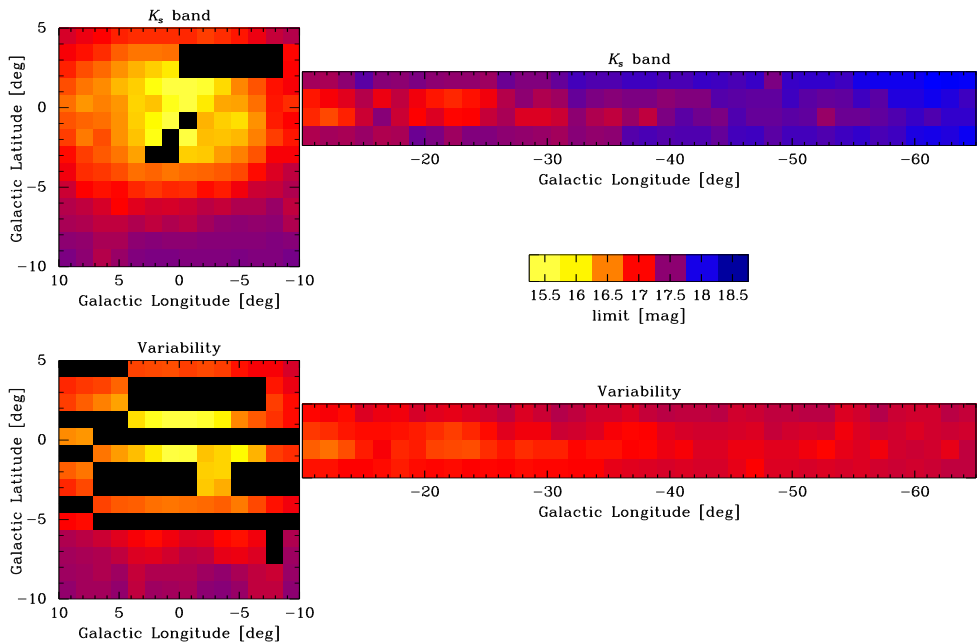}
\caption{The  $5\sigma$  magnitude  limit  of the  catalogues  in  the
  $K_{\rm s}$ band (top panel)  and for the variability campaign (also
  performed in the $K_{\rm s}$  band, bottom panel).  The colour scale
  is the same in both  maps.  Different strategies between the regular
  $K_{\rm s}$ observations and the variability campaign causes that in
  disk  fields the  variability data  show shallower  photometry.  The
  notation is similar to that present in Fig.~\ref{fig:maglim_zy}.}
\label{fig:maglim_kv}
\end{figure*}

While the whole area was observed  in the $JHK_{\rm s}$, 95\% of these
tiles satisfy  the stringent photometric and  image quality parameters
and are classified as completed.   In the $ZY$ bands the completion is
somewhat lower, with 59\% completed tiles.  The tiles completed in the
first   season   (until   October   26,   2010)   can   be   seen   in
Figs.~\ref{fig:maglim_zy},           \ref{fig:maglim_jh}           and
\ref{fig:maglim_kv},  respectively for  $ZY$, $JH$,  and  $K_{\rm s}$.
Figure~\ref{fig:maglim_kv} also  includes the tiles with  at least one
epoch  observed during  the variability  campaign in  the  $K_{\rm s}$
band.

In the first  season of the variability campaign 22  tiles in the disk
area had  5 $K_{\rm s}$  epochs taken, while  the majority had  one or
more  additional $K_{\rm  s}$ epochs  completed. The  completion rates
individually   for   the  bulge   and   disk   areas   are  given   in
Table~\ref{tab:completion}. Table~\ref{tab:tiles} states for each tile
whether the observations in a given filter are completed, and how many
additional $K_{\rm s}$ epochs were obtained until October 26, 2010.

\begin{figure}
\centering
\resizebox{\hsize}{!}{
\includegraphics[angle=0]{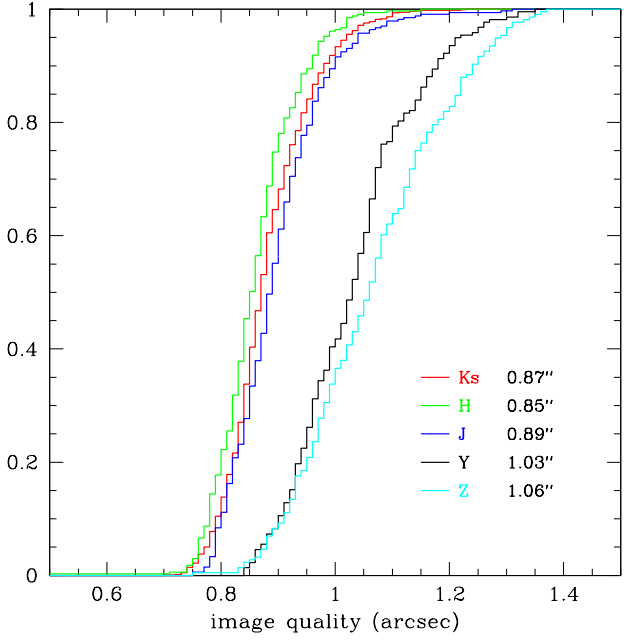}}
\resizebox{\hsize}{!}{
\includegraphics[angle=0]{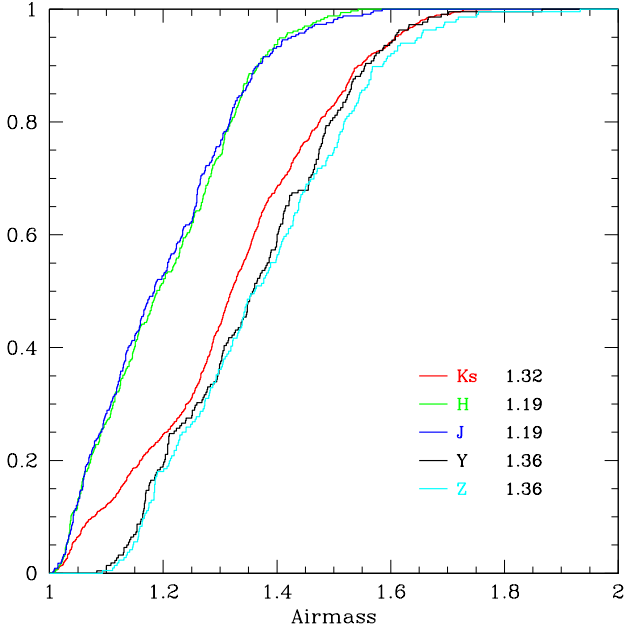}
}
\caption{Image  quality and airmass  cumulative distributions  for the
  VVV  observations obtained in  2010 in  the $ZYJHK_{s}$  filters are
  plotted  in the  top and  bottom panels,  respectively.   The median
  values of image quality and airmass for each filter are given in the
  legends. }
\label{fig:QCdistr}
\end{figure}

Figure~\ref{fig:QCdistr}  shows   cumulative  distributions  of  image
quality and  airmass for  the observed tiles  in the 2010  season. The
median image quality in the $J$, $H$ and $K_{\rm s}$ filters is better
than 0$\farcs$9 as measured on combined tile images, while it is close
to 1$\farcs$0 for the $Z$ and $Y$ filters.

\subsection{Observing strategy}
\label{sec:strategy}

The first observations collected for  the VVV Survey were taken during
the VISTA science  verification period in October 2009  when one field
in the Galactic  bulge at $\alpha$=18:02:58.872, $\delta=-$28:36:59.04
(J2000) was observed in the $ZYJHK_{\rm s}$ filters (called SV field).
In  addition to the  nearly-simultaneous multi-band  images of  the SV
field,  11  $K_{\rm  s}$  band   exposures  were  taken  to  test  the
variability observing strategy.  To  establish the necessary number of
exposures for proper sky subtraction, depending on crowding and number
of resolved objects in the field, three additional tiles were observed
in the  $K_{\rm s}$ band, two  bordering directly on the  SV field and
one  at an  offset position  $\approx$1$^\circ$ south  of  the science
target.  Based  on these early observations we  adjusted the observing
strategy,  which differs slightly  between the  bulge and  disk fields
(i.e., with  respect to exposure  time, number of  co-added exposures,
and combination of tiles for sky subtraction).

\begin{table}
\caption[]{Observing strategy and exposure times for VVV OBs.}
\label{tab:strategy}
\begin{center}
\begin{tabular}{lcccc}
\hline \hline
\noalign{\smallskip}
Area & Filter &  DIT$^{\,a}$  & NDIT$^{\,b}$  & Median exp. time \\
     &        & (s)   &       &  per pixel (s) \\
\noalign{\smallskip}
\hline
\noalign{\smallskip}
Bulge & $Z$, $Y$        &  10  &  1   & 40 \\
Bulge & $J$             &   6  &   2  &  48 \\
Bulge & $H$, $K_{\rm s}$  &   4  &   1  &  16 \\
Bulge & $K_{\rm s}$ (var) &   4  &   1  &  16 \\
\noalign{\smallskip}
\hline
\noalign{\smallskip}
Disk    & $Z$, $Y$            &  20   &  1   &  80 \\
Disk    & $J$, $H$, $K_{\rm s}$ &  10   &  2   &  80 \\
Disk    & $K_{\rm s}$ (var)     &   4   &  1   &  16 \\
\noalign{\smallskip}
\hline
\end{tabular}
$^{a}$\,Detector Integration Time\\
$^{b}$\,Number of DITs
\end{center}
\end{table}

Like all VISTA  observations the VVV Survey is  carried out in Service
Mode. The  basic observational unit  is the so-called  OB (observation
block).   The  multi-filter, single-epoch  OBs  have  been split  into
$JHK_{\rm s}$ and  $ZY$ OBs. The variability monitoring  OBs have only
single filter:  $K_{\rm s}$.   OBs for two,  three or four  tiles were
executed  back-to-back to ensure  that a  sufficient number  of offset
images were taken for each  filter to create a high quality background
sky frame.

The definition of the survey area  was done with the help of the SADT,
which  provides  the tile  centres,  guide  and  active optics  stars,
necessary for  efficient execution of  the survey OBs. Apart  from the
edges of the survey area, the  input to SADT is also the tile pattern,
which defines the large offsets  that fill in the inter-detector gaps,
as well as the size of  the smaller (jitter) offsets that are executed
at each  of the 6 pawprint  positions that together make  a tile.  All
VVV OBs  used the ``Tile6n''  pattern. In addition,  at each of  the 6
pawprint  positions  two  smaller   offsets  are  executed  using  the
``Jitter2u''  pattern.   This  means  that  in  total,  there  are  12
exposures  per filter,  but given  the  large offsets,  each pixel  is
covered  by at  least 4  exposures, except  for the  pixels  along the
y-edge  of the tile  (2 jitter  positions and  at least  2 pawprints).
These,  however,  have overlaps  with  adjacent  tiles. Therefore  the
complete  survey area  is  covered by  at  least 4  exposures in  each
filter.  Each  image is  a co-addition of  NDIT exposures  lasting DIT
seconds each.  The  total exposure times for bulge  and disk tiles are
given in Table~\ref{tab:strategy}.

\begin{figure}[ht]
\centering
\includegraphics[bb=1.5cm 8cm 15.5cm 15cm,angle=-90,scale=0.45]{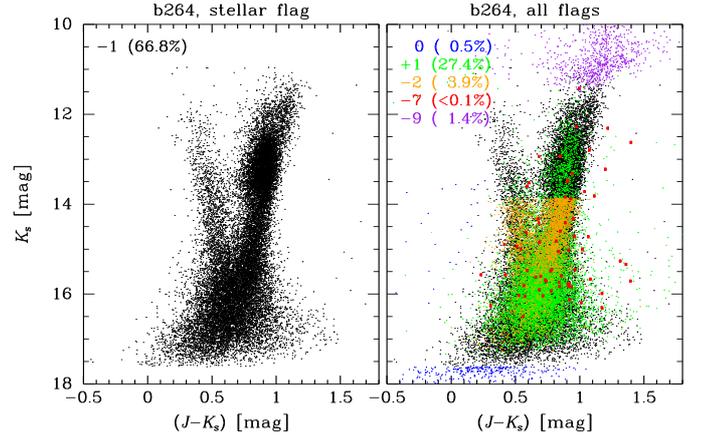}
\caption{$K_{\rm s}~vs.~(J - K_{\rm s})$ CMDs for a moderately crowded
  bulge  field (a  section of  tile  b264), showing  the high  quality
  sources with $-1$,  stellar flag (left panel), in  comparison to all
  other  flags  found  in  the  CASU catalogues  (right  panel).   The
  relative number of sources is given in the top left corner.}
\label{fig:flags}
\end{figure}

\begin{figure}[ht]
\centering
\includegraphics[bb=2cm 6cm 25cm 19cm,scale=0.5]{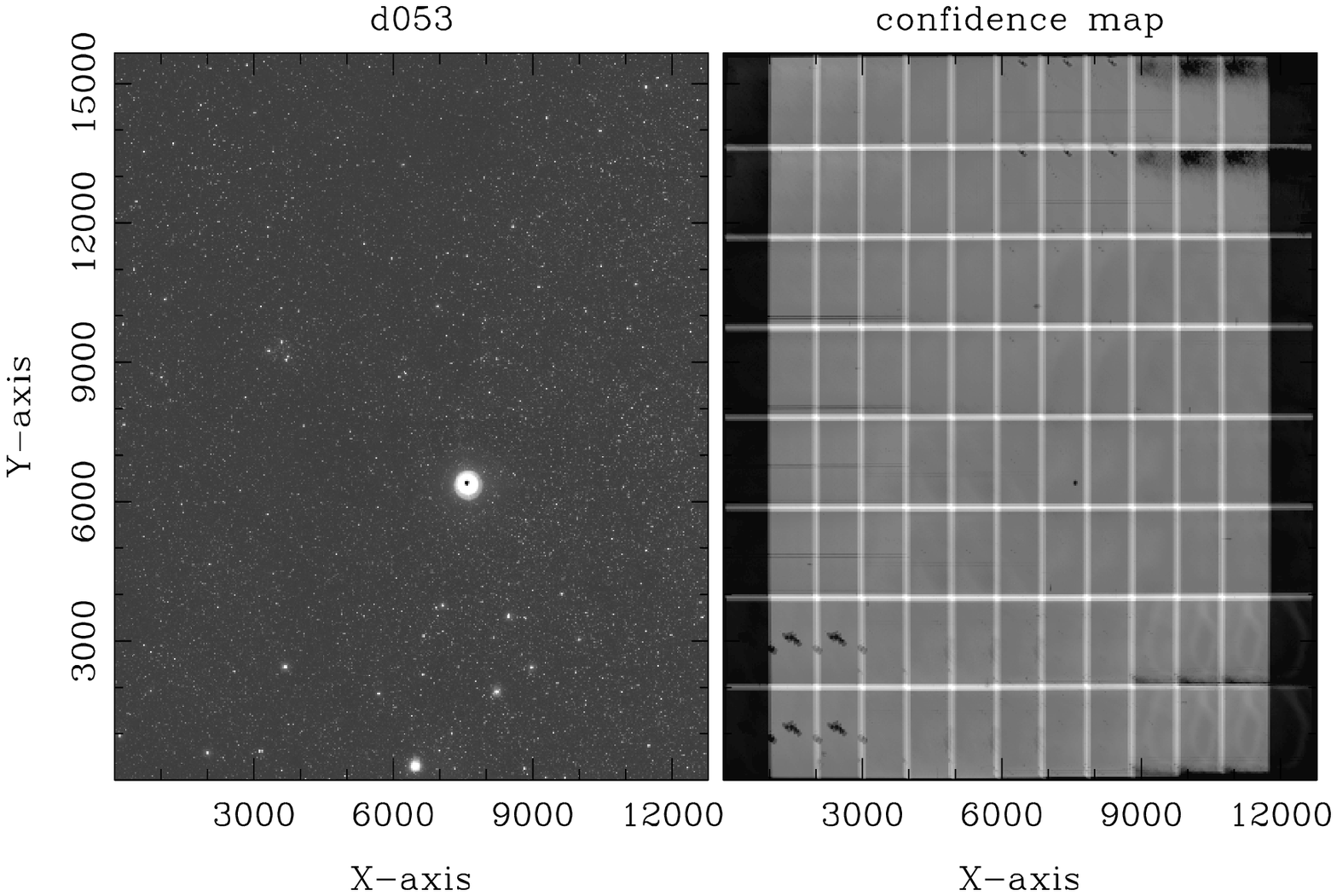}
\caption{The left  panel shows the disk  tile d053 in  the $K_{\rm s}$
  band, whereas its corresponding confidence map is shown in the right
  panel.  The brighter  regions in  the confidence  map have  a longer
  total exposure time, due to the combination of the 6 pawprints. This
  image  also includes 2  jitters at  each pawprint  position. Clearly
  visible in  the lower left corner  is a cluster of  bad pixels (Chip
  1).  The  six different positions  correspond to the  six pawprints,
  which are combined to fill the gaps between individual detectors.}
\label{fig:tile_confmap}
\end{figure}

\subsection{Data processing}
\label{sec:reduction}

VVV  observations are pipeline  processed within  the VISTA  Data Flow
System (VDFS)  pipeline at the  CASU \citep{2010ASPC..434...91L}.  The
processing is done  on a night-by-night basis, and  it consists of the
following data reduction steps executed in the order described.

The mean dark  current exposure, taken with the  same DIT (integration
time per exposure) and NDIT (number of coadds in the exposure) values,
is subtracted from each image.

A  linearity  correction is  applied  for  individual detectors  using
information on  the readout  time, exposure time  and the  reset image
time. A  ``reset'' exposure of  1.0s for every exposure  is subtracted
from  each  exposure within  the  data  acquisition  system, prior  to
writing the image to the disk.

The  flat-field correction  is done  by  dividing by  a mean  twilight
flat-field image  to remove small scale  quantum efficiency variations
and the  large scale vignetting profile  of the camera, as  well as to
normalize the gain of each detector to a common median value.

The  sky  background  correction   removes  the  large  scale  spatial
background  emission.   Tests   made  with  the  science  verification
observations showed that 12  exposures (6 pawprint $\times$ 2 jitters)
taken for each  VVV tile do not yield good  sky subtraction because of
severely crowded fields, leaving  ``holes'' at the positions of bright
stars  or   very  crowded   regions  with  many   overlapping  stellar
PSFs. Therefore for VVV a  sky background map is produced by combining
all exposures for a given filter taken within several concatenated OBs
(tiles).  Due to  variability of the near-IR sky, the  need to take at
least 24 images with the same filter within $\sim 30$~min poses limits
to the exposure times for individual tiles and the number of different
filters that can be included in each OB.

A ``destriping correction'' is  performed by subtracting the low level
horizontal stripe pattern introduced by the readout electronics of the
VIRCAM detectors.

Jitter  stacking is  performed to  align two  slightly  shifted images
taken at a given pawprint position, combining them into a single image
for each  pawprint.  The  shifts are computed  using the  positions of
many hundreds of stars detected in all images.

Object  detection  is  performed  for  each  stacked  pawprint  image.
Positions, fluxes  measured in  several apertures of  different sizes,
and some  shape measurements are  written in the source  catalogue.  A
flag indicates the most  probable morphological classification, and in
particular we  note that ``$-1$''  is used to denote  stellar objects,
``$-2$'' borderline stellar, ``$0$'' is noise, and ``$+1$' is used for
non-stellar  objects.   There are  also  objects  with flag  ``$-7$'',
denoting sources containing bad pixels,  and the flag ``$-9$'' is used
for  saturated  stars.   These  flags  are  derived  mainly  based  on
curve-of-growth  analysis  of  the  flux  \citep{2004SPIE.5493..411I}.
Figure~\ref{fig:flags}  shows colour-magnitude  diagrams (CMDs)  for a
moderately crowded  bulge field (b264), comparing  the distribution of
the high quality sources, with  all the other flags.  Relative numbers
for each  flag are  also indicated  in the figure.   In addition  to a
catalogue  with extracted sources  a confidence  map is  also computed
(see Fig.~\ref{fig:tile_confmap}).

The six stacked  pawprint images are combined into  a single deep tile
and  the   catalogue  extraction   step  is  repeated.    Tile  images
contiguously   sample   about   $1.5   \times  1.1$~sq.~deg   on   the
sky. Confidence maps are  computed for each tile, relaying information
on  the   different  exposure  times   for  pixels  across   the  tile
(Fig.~\ref{fig:tile_confmap}). The  exposure times depend  on the size
of the  jitter offsets and the  number of exposures  that are combined
into a tile.  The confidence  maps clearly show the areas of detectors
affected by bad  pixels, such as the large patch  in detector 1 (lower
left  corner  in  the  right  panel  of  Fig.~\ref{fig:tile_confmap}),
several rows in detector 4 (lower right corner) and the upper third of
detector  16 (upper  right  corner). After  stacking several  dithered
exposures  in  the  final tile,  most  of  these  bad pixels  are  not
noticeable (left panel  of Fig.~\ref{fig:tile_confmap}), but the large
bad  area  on  detector  16  has larger  errors  in  the  illumination
correction map.  The  sensitivity of that upper third  of the detector
is much worse at shorter wavelengths, and some offsets with respect to
2MASS calibration for  the whole tile have been found  in $Z$, $Y$ and
$J$ bands. Offsets  at longer wavelengths, in the  $H$ and $K_{\rm s}$
bands, are within the calibration errors.

The  first  year data  release  of  the  tile images,  catalogues  and
confidence maps  described here is version  1.1.  Detailed information
about  this  version  of  VISTA  data  products as  well  as  a  brief
description  of  all  issues  encountered during  data  processing  is
available     on     the      CASU     web     page     \footnote{{\tt
    http://casu.ast.cam.ac.uk/surveys-projects/vista      /technical}}.

\subsubsection{Generation of multi-band catalogues}

Currently the  version 1.1 single  band tile catalogues from  CASU are
matched   by  the   VVV  team   members  using   the   STILTS  package
\citep{2006ASPC..351..666T} and a KD-Tree based algorithm (Gurovich et
al. 2011,  in prep.)  that  uses the Cross  et al.  (2011,  in prep.)
implemented source matching method for  the VSA data.  The matching is
done using astrometrically corrected  tiles and catalogues, allowing a
1$\farcs$0 offset between  point sources  to be  considered  a match.
Nevertheless, tests have  shown that over 90\% of  the stellar sources
are  matched  within less  than  $0\farcs  5$,  as expected  from  the
astrometric       accuracy      of      the       catalogues      (see
Sec.~\ref{sec:astrometry}).

After  matching  the single  band  catalogues,  most  of the  spurious
detections  around  bright  stars  are  rejected.   Unfortunately  the
ellipticity sometimes  varies from image to image;  which then results
in  rejection of  some sources  which  are only  classified as  ``bona
fide'' stars in selected  filters, but appear slightly more elongated,
and  are thus  rejected as  non-stellar,  in other  filters.  A  final
multi-band catalogue  contains $\sim75-85\%$  of the sources  found in
the catalogue with the least  number of sources used for the matching,
in general  the $K_{\rm s}$  band, for low to  intermediate extinction
regions.   Close to the  Galactic plane  high extinction  affects more
severely the  source detection at short wavelengths,  hence the source
density is  limited by the  detection rate in  the $Y$ and  $Z$ bands.
The comparison between the different tiles is hampered not only by the
different median extinction values,  but also by the strong extinction
variation within any  given tile. As an example,  we divided two tiles
(b305     and     d003)    into     0.25$^\circ$~$\times$~0.25$^\circ$
sub-sections.  Median  extinction values  A$_{V}$  and their  standard
deviations are given in Table \ref{tab:extinct}.

\begin{table}

\caption[width=\textwidth]{Galactic  extinction  values  (median)  for
  tiles b305  and d003, assuming  $A_{V} =3.1 \times E(B-V)$ and based
  on      the      prescriptions     by      (1)~\citet{1998ApJ...500..525S},
  (2)~\citet{2003A&A...409..205D},   (3)~\citet{2005AJ....130..659A},
  (4)~\citet{2006A&A...453..635M},    (5)~\citet{2005A&A...432L..67F},
  (6)~\citet{2005PASJ...57S...1D}}
\label{tab:extinct}
\begin{center}
\begin{tabular}{lccc}
\hline
\hline
\noalign{\smallskip}
Tile    & Reference            & $A_{V}$ (median) & $\sigma(A_{V})$ \\
\noalign{\smallskip}
\hline
\noalign{\smallskip}
b305    & $(1)$          &  4.04            &  1.54              \\
        & $(2)$          &  4.81            &  1.81              \\
        & $(3)$          &  3.11            &  0.67              \\
        & $(4)$          &  3.95            &  1.22              \\
        & $(5)$          &  4.21            &  1.17              \\
\hline
\noalign{\smallskip}
d003    & $(1)$          &  5.73            &  2.05              \\
        & $(2)$          &  5.67            &  0.89              \\
        & $(3)$          &  3.25            &  0.91              \\
        & $(4)$          &  5.12            &  1.99              \\
        & $(5)$          &  2.66            &  0.66              \\
        & $(6)$          &  2.72            &  0.23             \\
\noalign{\smallskip}
\hline
\end{tabular}
\end{center}
\end{table}

Multi-band  catalogues corresponding to  the DR1  data have  also been
generated by  the VISTA Science  Archive (VSA). While the  single band
DR1  data  were  already   delivered  through  the  ESO  Archive,  the
multi-band catalogues  will be publicly available in  due time through
the     ESO     Archive      and     VSA     web     page\footnote{\tt
  http://horus.roe.ac.uk/vsa/index.html}.    VSA  will   also  provide
multi-epoch  catalogues  for  variable  sources.   Part  of  the  data
reduction and  storage are performed  using the Geryon cluster  at the
Center    for     Astro-Engineering    at    Universidad    Cat\'olica
(AIUC)\footnote{\tt http://www.aiuc.puc.cl/}.

\begin{figure}[ht]
\centering
\includegraphics[bb=1cm 7cm 25cm 26cm,scale=0.6]{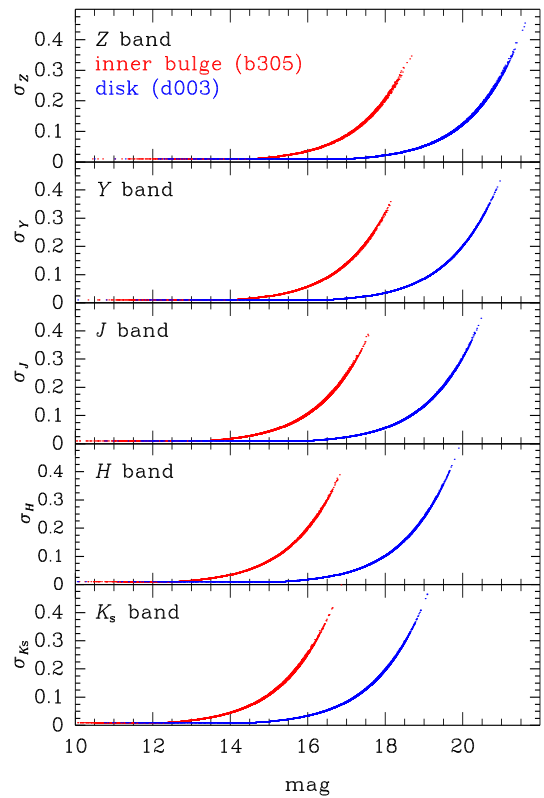}
\caption{Photometric  error  as  a  function  of  magnitude  for  each
  passband, $ZYJHK_{\rm  s}$, for a  representative field in  the disk
  (d003) and  for a crowded bulge  field close to  the Galactic Centre
  (b305).   We note  that different  exposure times  between  disk and
  bulge areas  also contribute  to the shallower  curve for  the bulge
  data.}
\label{fig:mag_sigma}
\end{figure}

\section{Photometry}
\label{sec:photometry}

Photometric calibration of stacked  pawprint images and tile images is
done using  numerous detected 2MASS stars.   The calibration procedure
follows  closely that  of WFCAM  \citep{2009MNRAS.394..675H}. Internal
photometric accuracy  is of the order  of $\pm 2$\%, and  for the $J$,
$H$ and $K_{\rm s}$ bands a  similar accuracy as for 2MASS is achieved
for most of the survey  area.  In particularly high extinction regions
and for the $Z$ and $Y$ filters the photometric calibration errors are
somewhat larger.

Figure~\ref{fig:mag_sigma} shows the  photometric errors as a function
of the magnitude computed by  the CASU pipeline for the five passbands
in a  typical disk field  (d003) as  well as in  an extreme case  of a
crowded  bulge field close  to the  Galactic Centre  (b305). Different
extinction      \citep[$E(B-V)_{\rm      d003}=1.77$,     $E(B-V)_{\rm
    b305}=1.37$~mag;][]{1998ApJ...500..525S}  and crowding  levels, as
well  as distinct observing  strategies between  disk and  bulge areas
(see Section~\ref{sec:strategy}),  contribute to the  shallower curves
seen for  the bulge data in  all five passbands.  The  accuracy in the
photometry can  also be checked using the  overlapping regions between
the  tiles.  Figure~\ref{fig:astrometry_VVVoverlap}  shows in  the top
panels the  $K_{\rm s}$ photometry for stellar  sources (also computed
by the  VDFS pipeline at CASU)  in the overlapping  area between tiles
b305 and  b291, and  between d003 and  d041.  The  $5\sigma$ magnitude
limits reached  for each  passband, for both  bulge and disk  tiles as
computed  by  the pipeline,  are  shown in  Figs.~\ref{fig:maglim_zy},
\ref{fig:maglim_jh} and \ref{fig:maglim_kv}.

\subsection{Saturation}

VIRCAM   detectors  saturate  at   different  levels,   mostly  around
$33000-35000$~ADU.    Detector  \#5  has   the  smallest   well  depth
(saturation at  $\sim 24000$~ADU), but  detector \#13 has  the largest
non-linearity with of  $\sim 10$\% at 10000 ADU.   This, combined with
the rather bright near-IR sky results in a restricted dynamic range in
the photometry.

Prior to  reaching saturation  VISTA detectors have  nonlinear regime.
Although  linearity  correction  is  performed in  the  pipeline  data
reduction, based  on observations of  an illuminated dome  screen, for
the  stars  close  to  the  saturation limit  there  are  still  quite
significant  deviations   with  respect  to  2MASS   due  to  residual
non-linearity.    This   can  be   seen   clearly   in   Fig.   2   of
\citet{2011A&A...534A...3G}  and  also  in  the VMC  survey  paper  by
\citet{2011A&A...527A.116C}.

\begin{figure}[ht]
\centering
\includegraphics[bb=1.2cm 9cm 25cm 25cm,scale=0.55]{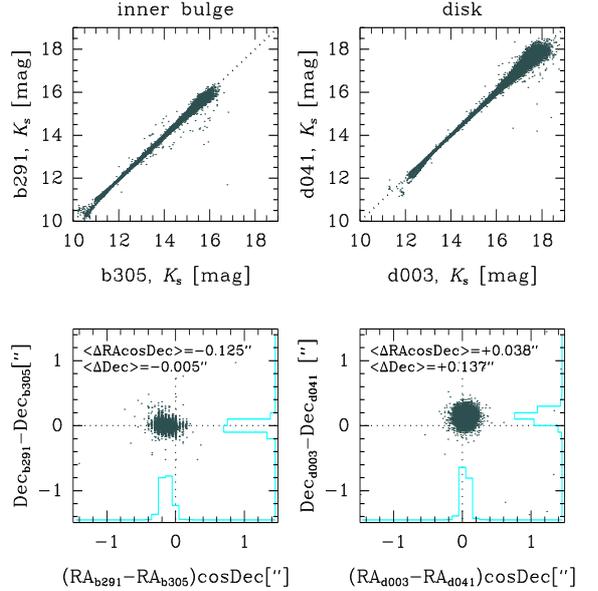}
\caption{Top  panels:  Photometry in  the  $K_{\rm  s}$  band for  the
  overlapping region between tiles b291  and b305 (left), and d003 and
  d041 (right).  Only stellar sources were used in these plots. Bottom
  panels: Astrometric  accuracy for the same  overlapping regions. The
  mean     values     for    $\Delta\alpha$$\times$cos$\delta$     and
  $\Delta$$\delta$  are  shown  in  the  top  left  corner.   Counting
  histograms for the distribution are also shown for both axes.}
\label{fig:astrometry_VVVoverlap}
\end{figure}

The overlapping  regions also  help us to  check the linearity  of the
photometry   at  the  saturation   limit.   In   the  top   panels  of
Fig.~\ref{fig:astrometry_VVVoverlap},  for both  bulge and  disk data,
part  of  the  brightest   stars  slightly  scatter  from  the  linear
distribution close  to the saturation level, even  taking into account
that  these are  ``$-1$'' (stellar)  sources in  all VDFS  single band
catalogues from  CASU.  Some saturated  stars are also present  in the
CASU catalogues  (flagged as ``$-9$'',  see Fig.~\ref{fig:flags}), but
these are not present in Fig.~\ref{fig:astrometry_VVVoverlap}.

\subsection{Photometric Completeness}
\label{sec:complet}

Although the main goal of this paper is to describe the content of the
first data release  for the VVV Survey we  want to demonstrate briefly
how  strongly the completeness  of the  source catalogues  provided by
CASU depends on the location of the selected area within the Survey.

To do  so we  carried out artificial  star experiments,  comparing the
detection rate for artificial stars (AS) added to the images using the
CASU       source      detection       package       {\it      imcore}
\citep[][]{2004SPIE.5493..411I}.  We selected two different tiles from
the bulge area, namely b204 and b314, representing different levels of
crowding. From these tiles we cut a small 2000 $\times$ 2000 pixel$^2$
area  ($11\farcm3 \times 11\farcm3$)  for the  completeness tests.
The   two  fields   were   centred  on   $\alpha_{2000}$=18:12:13.173,
$\delta_{2000}$=-38:07:49.220                 for                 b204
($l,b=354.72^\circ,-9.37^\circ$),   and  $\alpha_{2000}$=17:29:28.920,
$\delta_{2000}$=            -36:00:09.800           for           b314
($l,b=352.21^\circ,-0.92^\circ$),   respectively.    The  field   b204
represents a  less crowded section  of the survey area,  whereas b314,
close to the  Galactic Centre, is one of the  most crowded fields. The
$K_{s}$  band  source  catalogues   for  the  complete  tile  and  the
respective   sub-sections    contain   $482,004/11,584$   (b204)   and
$1,137,615/23,691$ (b314) sources.  We point out that in a preliminary
test we applied the source detection package to the complete b204 tile
and confirmed  the number  of sources obtained  by the  CASU pipeline,
hence validating the source detection in the AS experiments.

In our completeness tests we added  5000 AS to the original images for
16 individual magnitude intervals, each 0.2~mag wide. The positions of
these stars were  chosen randomly, but the same  positions and $K_{s}$
magnitudes were used  for both fields.  To create the  AS we derived a
single point  spread function (PSF),  using isolated stars  within the
corresponding image.  We leave  the completeness test using a variable
PSF for  later.  The  AS images were  constructed using the  IRAF task
{\it addstar}  \footnote{IRAF is  distributed by the  National Optical
  Astronomy Observatories,  which are  operated by the  Association of
  Universities  for  Research in  Astronomy,  Inc., under  cooperative
  agreement with the US National Science Foundation.}.

\begin{figure}[ht]
\centering 
\includegraphics[bb=6.5cm 13cm 15.5cm 26cm,scale=0.54]{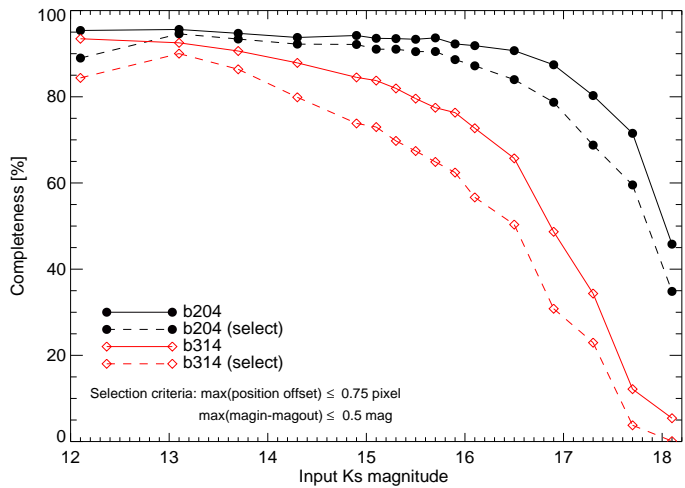}
\caption{$K_{\rm s}$  band completeness test  for two $2000\times2000$
  pixels$^2$ ($11\farcm3 \times 11\farcm3$) fields from the bulge area
  of the  VVV Survey.   The fields were  selected to represent  a less
  crowded  region (tile  b204)  as well  as  one of  the most  crowded
  regions  near  the  Galactic   Centre  (tile  b314).   Solid  points
  correspond  to  b204,  whereas  open diamonds  represent  b314.   In
  addition we  apply selection criteria  to define a  confirmed source
  detection in our AS experiments (see text for details).  Solid lines
  show the  completeness fraction  if a source  was detected  within a
  radius $\le$ 1 pixel of its original position.  Reducing this radius
  to 0.5 pixel  and requiring the original $K_{\rm  s}$ band magnitude
  to be  recovered within 0.5  mag results in the  completeness curves
  shown with dashed lines. }
\label{fig:completeness}
\end{figure}

Using {\it  imcore} we  created source catalogues  for each of  the AS
images, which  we call  ``output'' and compared  them to  the original
source list ``input''. The ratio  of recovered AS to the original 5000
stars for a  magnitude range between $12.0\leq K_{\rm s}\leq18.2$~mag
is  shown in  Fig.~\ref{fig:completeness}.  In  the first  test (solid
lines)  we consider  a positive  AS detection  if a  source  was found
within 1 pixel of the  inserted position.  However, in addition to the
simple  detection  we  also  need  to know  with  which  accuracy  the
``input'' magnitude is recovered.  Based  on the AS experiments of the
``ACS              Globular              Cluster              Survey''
\citep[][]{2007AJ....133.1658S,2008AJ....135.2055A} we first allowed a
0.75~mag offset  between input and output magnitude,  which we finally
reduced  to 0.5~mag.   This  is based  on  the fact  that the  stellar
densities in  both our  fields are assumed  to be  significantly lower
than that within the central region of a globular cluster, targeted by
the  ACS Globular Cluster  Survey.  The  completeness curve  for those
tests is shown in Fig.~\ref{fig:completeness} by dashed lines.

Based on  the different completeness  pattern we find that  the source
detection  efficiency reaches  50$\%$  for stars  with $17.8  \lesssim
K_{\rm s} \lesssim 18.1$~mag in  field b204, and $16.4 \lesssim K_{\rm
  s} \lesssim 16.9$~mag in  field b314, respectively, depending on the
restrictions applied.  We also  note that the completeness is somewhat
lower for  the brightest stars (12.0  to 12.2~mag) in  both fields.  A
possible  explanation  may  be  that  those stars  are  close  to  the
saturation limit and the photometric accuracy is therefore reduced.

\section{Comparison with 2MASS photometry}
\label{sec:2mass}

\begin{figure}[ht]
\centering 
\includegraphics[bb=6.cm 12cm 15.5cm 26cm,scale=0.55]{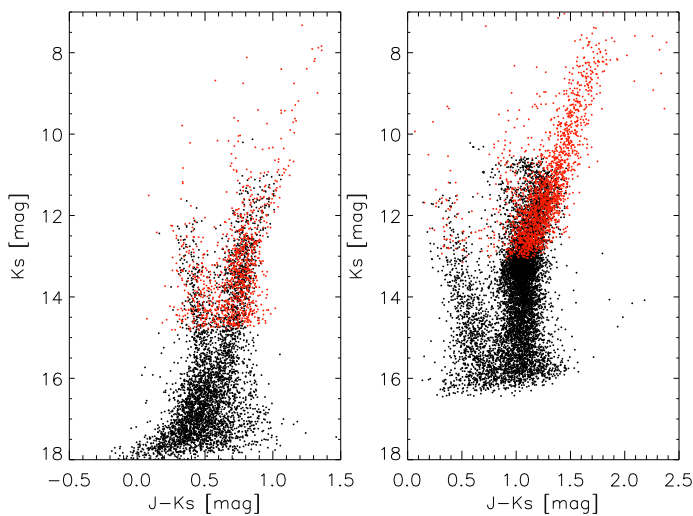}
\caption{CMDs comparing VVV (black) and 2MASS data (red) for the bulge
  area.  The left-hand  panel shows a field in  the outer bulge (b235)
  while  the right-hand  panel shows  one  of the  most crowded  bulge
  fields (b305), close to the Galactic Centre.}
\label{fig:2mass_vvv}
\end{figure}

The VVV observations are approximately 4 magnitudes deeper than 2MASS.
In addition,  a very important  factor is the excellent  image quality
(with seeing  $0\farcs9-1\farcs0$) in the  entire VVV Survey  area for
the multi-band, single epoch observations. This allows us to reach the
red      clump     magnitude      across     the      entire     bulge
\citep[][]{2011A&A...534A...3G,2011AJ....142...76S},  and therefore to
study the stellar populations and the structure of the inner Galaxy to
an unprecedented level of detail, as for example, identification of RR
Lyrae and derivation of accurate distances. Figure~\ref{fig:2mass_vvv}
shows a CMD for $10' \times 10'$ regions in the inner bulge (b305) and
the outer bulge  (b235) of VVV stellar sources  compared to 2MASS CMDs
of the same  regions.  As the VVV photometry is  much deeper it allows
us not only to trace the red clump even in the most extincted regions,
but  also  to study  the  stellar  populations  behind the  bulge.   A
detailed study  of the colour transformations between  VISTA and 2MASS
photometric systems for  VVV disk fields will be  presented by Soto et
al. (2011, in prep.).

\begin{figure}[ht]
\centering
\includegraphics[bb=1.3cm 3cm 25cm 25cm,scale=0.49]{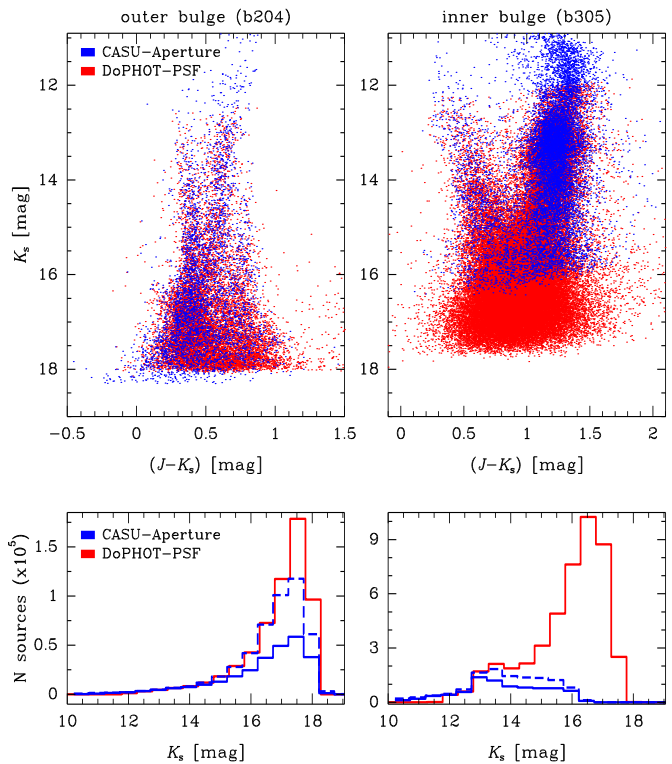}
\caption{The top-left  panel shows a $K_{\rm s}~vs.~(J-K_{\rm s})$
  CMD  for  a  less crowded  bulge  field  (a  section of  tile  b204)
  comparing  the  $5\sigma$  magnitude  limit  of  the  CASU  aperture
  photometry  with  PSF  photometry  (also  with  $5\sigma$  magnitude
  limit).   The bottom-left  panel  shows histograms  for all  sources
  (dashed line)  and only  the stellar objects  (solid line)  found in
  tile  b204,  also  in  comparison  with  the  PSF  photometry.   The
  right-hand panels show the same  analysis for a highly crowded field
  close to the Galactic Centre (tile b305).}
\label{fig:aper_psf2}
\end{figure}

\begin{figure}[ht]
\centering
\includegraphics[bb=2.1cm 12cm 25cm 26.3cm,scale=0.55]{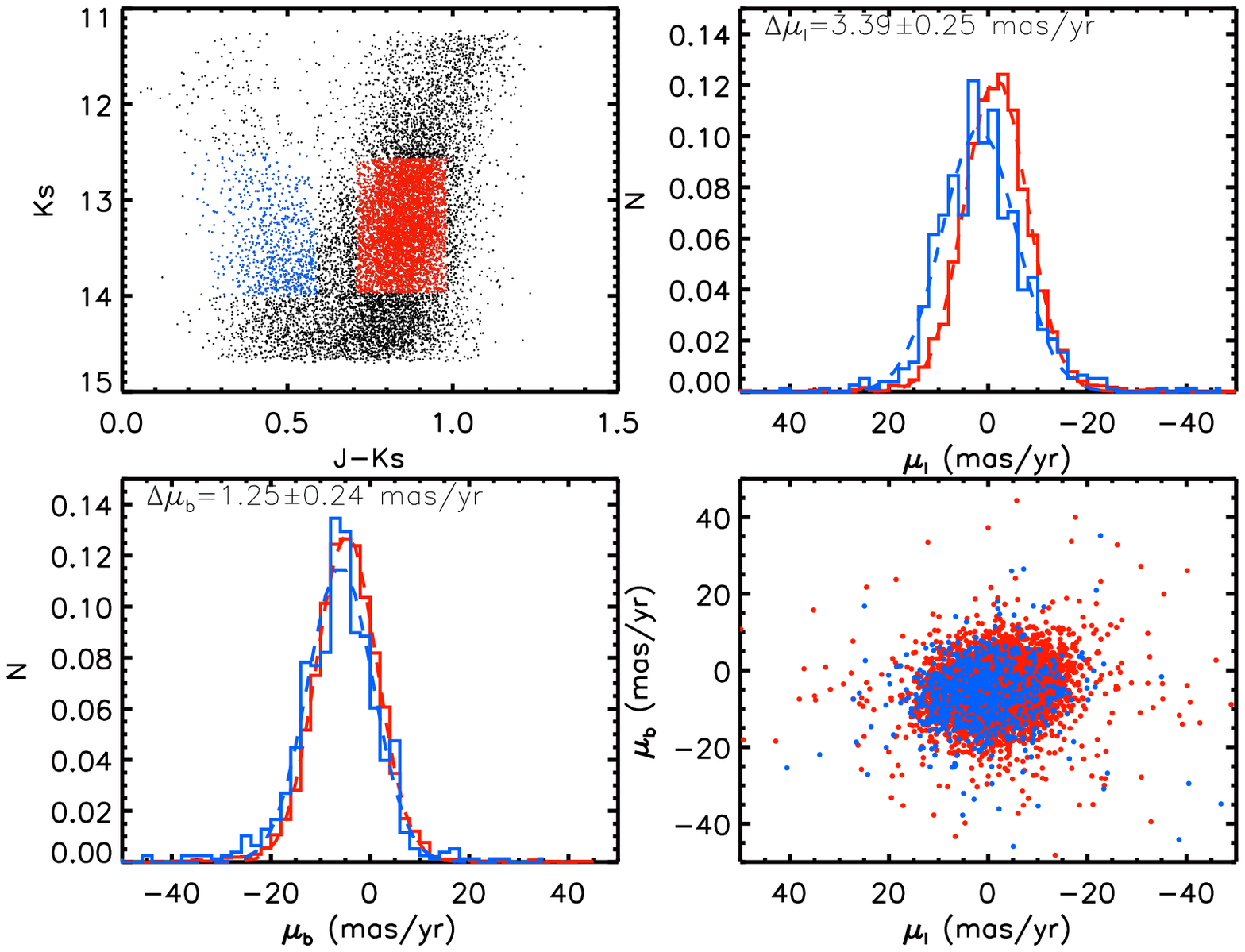}
\caption{Top left  panel: $K_{s}~vs.~(J-K_{\rm s})$ CMD for  a $10
  \times 10$ arcmin$^{2}$ region in  tile b249 for VVV stellar sources
  matched with 2MASS.   The CMD is centred in the  region of the bulge
  red clump (red) and the disk main sequence (blue).  Top right panel:
  histogram  of  the  longitudinal  proper motion  for  both  selected
  regions     in     the     CMD.      The     mean     values     for
  $\Delta\alpha$$\times$cos$\delta$ and ${\rm \Delta\delta}$ are shown
  in the top left corner.  Bottom  left panel: same as top right panel
  for the latitudinal proper  motion distribution. Bottom right panel:
  the $\mu_b~vs.~\mu_l$ distribution in units of mas~yr$^{-1}$.}
\label{fig:astrometry_VVV_2MASS}
\end{figure}

\section{VVV DR1 catalogues vs. PSF photometry}
\label{sec:psf}

The photometric catalogues published in this, as well as any following
VVV data release,  are based on aperture photometry  only, computed by
the      CASU     pipeline      on      individual     tiles      (see
Section~\ref{sec:reduction}).  However, for very crowded fields (e.g.,
the  innermost  Galactic  Centre  or  the central  region  of  stellar
clusters), more  complete and deeper  photometry can be  obtained with
PSF photometry.   In this  section we present  tests performed  by VVV
team members using PSF photometry on the VVV images.  The results show
that the  PSF fitting can reach  up to $1.5$~mag  deeper than aperture
photometry,  detecting up  to twice  more sources  for  highly crowded
fields,  where  aperture  photometry   is  known  to  be  inefficient,
particularly for  faint sources.  We  emphasize that the PSF  data are
not part of the VVV DR1.

The  tests were performed  using the  {\it apermag3}  aperture fluxes,
that are used in the CASU catalogues as well as in the VSA database as
the default values to represent the flux for all images.  However, the
CASU pipeline  measures positions and fluxes  for different concentric
apertures  designed to  adequately sample  the curve-of-growth  of the
majority of images.  The {\it apermag1} has 1\arcsec~diameter and each
successive aperture  increases by a  factor of $\sqrt 2$  in diameter.
For  highly crowded  fields  the aperture  photometry  using the  {\it
  apermag3} can  simultaneously fit multiple  overlapping sources, but
it  does not perform  any subtraction  in order  to check  for fainter
sources hidden underneath.   In this case the {\it  apermag1} and {\it
  apermag2} can be more suitable than {\it apermag3}, even taking into
account that small apertures have more uncertain aperture corrections,
especially in poor seeing conditions.

Figure~\ref{fig:aper_psf2}  shows in  the top-left  panel  the $K_{\rm
  s}~vs.~(J-K_{\rm s})$  CMD for a field with  relatively low crowding
and extinction (a section of tile b204), comparing aperture photometry
performed   by  CASU   with  PSF   photometry  obtained   with  DoPHOT
\citep{1993PASP..105.1342S}  for  the   same  field,  using  $5\sigma$
magnitude limits in  both cases. All structures are  correctly seen in
both  cases, and  both  aperture and  PSF  photometries reach  $K_{\rm
  s}\sim18.0$~mag.

The bottom-left panel  of Fig.~\ref{fig:aper_psf2} shows the magnitude
distribution of all sources found  in the CASU catalogue (dashed line)
as well  as of  stellar sources only  (solid line).  Even  taking into
account that aperture photometry  contain some brighter stars that are
not present in  the PSF photometry, the total  number of sources found
by PSF photometry in the $K_{\rm  s}$ band image of b204 is $589,187$,
in comparison to $482,004$  (all sources) and $273,550$ (stellar flag)
sources present in the CASU catalogue.

For comparison, the  same analysis is performed in  the right panel of
Fig.~\ref{fig:aper_psf2}  for the  inner bulge  field b305,  where the
effects of  crowding and  extinction are significant.   PSF photometry
allows us  not only  to resolve  the high density  areas, but  it also
reaches $\sim1.5$~mag deeper, i.e., $K_{\rm s}\sim17.5$~mag, while the
aperture  photometry  is  limited  to  $K_{\rm  s}\sim16.0$~mag.   The
bottom-right panel of Fig.~\ref{fig:aper_psf2}  shows how much the PSF
photometry  exceeds the  aperture photometry  for faint  sources.  The
total number of sources found by  PSF photometry for field b305 in the
$K_{\rm  s}$ band is  $4,601,529$ in  comparison to  $1,201,557$, (all
flags)  and  $818,706$ (stellar  only)  sources  present  in the  CASU
catalogue.

Tests performed by  VVV team members show that  the PSF photometry can
be   even  deeper   using   a  DAOPHOT-ALLFRAME   suite  of   routines
\citep{1994PASP..106..250S}  customized  for the  VVV  data (Mauro  et
al. 2011, in prep.).

\section{Astrometry}
\label{sec:astrometry}

The  native VISTA  WCS  distortion  model for  pawprints  is based  on
Zenith-Pole-North (ZPN) projection and  is available in image headers.
The distortions are radial and are well described by:

\begin{equation}
r' = k_1 \times r + k_3 \times r^{3} + k_5 \times r^5,
\end{equation} with $k_1=0.3413$~arcsec/pix being the plate scale at 
the  centre,  and  $k_3/k_1  =  44$,  $k_5/k_1=$~10,300  are  distortion
coefficients  in angular  units of  radians.  Higher  order  terms are
negligible.

The median  WCS $rms$ is $\sim  70$~mas and is dominated  by the 2MASS
astrometric errors.  We note that in both coadded pawprints as well as
complete  tiles the  pixels have  been resampled  to a  common spatial
scale and  the astrometric distortions have been  removed according to
equation (1).  The resampling was also required in the pawprints since
the  jitter  offsets are  sufficiently  large  (i.e., 15$\farcs$0)  to
affect the  coadding (see also  Table \ref{tab:strategy}), due  to the
size  of  the  distortions.   Another  way to  evaluate  the  internal
astrometric  accuracy  is to  use  overlaps  between different  tiles.
Hundreds of  stars are detected  independently on two  adjacent tiles.
Fig.~\ref{fig:astrometry_VVVoverlap}  shows  in  the  top  panels  the
photometry in $K_{\rm s}$ for overlapping regions between tiles in the
inner bulge (b291 and b305) and disk (d003 and d041) areas.  Different
observing strategies, in particular  exposure times (see Table 3), for
the disk  and bulge areas, as  well as the  high background brightness
due to unresolved stars in  the inner bulge, make the bulge photometry
shallower than that  in the disk region.  The  astrometric accuracy is
shown in the  bottom panels of Fig.~\ref{fig:astrometry_VVVoverlap} in
terms       of        the       distribution       $\Delta\delta$~{\it
  vs.}~$\Delta\alpha$$\times$cos$\delta$.   Typical   values  for  the
astrometric  accuracy  are  $\sim25$~mas  for a  $K_{\rm  s}=15.0$~mag
source  and  $\sim175$~mas  for  $K_{\rm  s}=18.0$~mag.   Despite  the
shorter time  baseline of  the VVV Survey,  the typical  proper motion
measurements      should      reach      the     accuracy      between
$\sim7$~mas~yr$^{-1}$~($K_{\rm             s}=15.0$~mag)            to
$\sim15$~mas~yr$^{-1}$~($K_{\rm s}=18.0$~mag) after  the five years of
the VVV campaign.

A test for  such measurements can already be  achieved using 2MASS and
VVV datasets which provide a time baseline of 11 years. Cross-matching
of sources  between VVV and 2MASS  was performed for a  $10 \times 10$
arcmin$^2$ field  in tile  b249.  The astrometric  differences between
stellar sources in  these two catalogues can be  used to derive proper
motions    of   stars    in    terms   of    $\mu_l$   and    $\mu_b$.
\cite{2008ApJ...684.1110C}, based on two epochs of HST imaging, showed
that foreground  disk stars observed in  the colour-magnitude diagrams
towards  the bulge  can be  separated based  on Gaussian  fits  to the
distribution  of  their  proper   motions  with  mean  differences  of
$(\Delta\mu_l,\Delta\mu_b)=(3.22\pm0.15,0.81\pm0.13)$~mas~yr$^{-1}$.

Figure~\ref{fig:astrometry_VVV_2MASS} shows the  selection of stars in
the CMD for  matched stars between VVV and 2MASS  which belong to disk
and bulge  populations together with  the distribution of  $\mu_l$ and
$\mu_b$ for each selection.  Although the distributions are wider than
those presented in the HST analysis of \cite{2008ApJ...684.1110C}, the
mean differences  between proper motions  of disk and bulge  are still
evident    based    on    these    data   showing    differences    of
$(\Delta\mu_l,\Delta\mu_b)=(3.39\pm0.25,1.25\pm0.24)$~mas~yr$^{-1}$,
in good agreement with those of \cite{2008ApJ...684.1110C}.

\section{VVV Source maps}
\label{sec:maps}

\begin{figure*}[ht]
\centering          
\includegraphics[bb=0cm 4.0cm 20cm 22cm,angle=-90,scale=0.50]{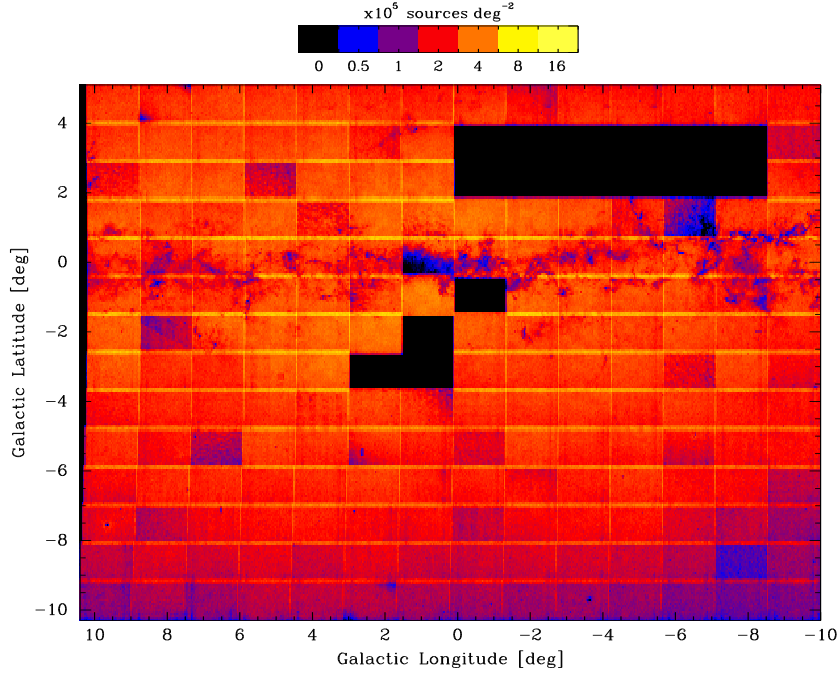}
\caption{Density map in logarithmic  scale showing the VVV bulge area.
  The map was made using all the stellar point sources detected in the
  $J$, $H$ and $K_{\rm s}$  1.1 CASU catalogues.  Crowded areas appear
  in yellow, while  less populated regions as well  as high extinction
  areas are shown  in blue. The overlapping regions  between the tiles
  are  highlighted  since  the   point  sources  are  detected  twice,
  therefore generating the grid  pattern, that indicates also the size
  of    overlap     regions.     The    density,     in    units    of
  10$^5$~sources~deg$^{-2}$, is indicated in the horizontal bar at the
  top.}
\label{fig:bulge}
\end{figure*}

\begin{figure*}[ht]
\centering      
\includegraphics[bb=12cm 3.2cm 19cm 22cm,angle=-90,scale=0.72]{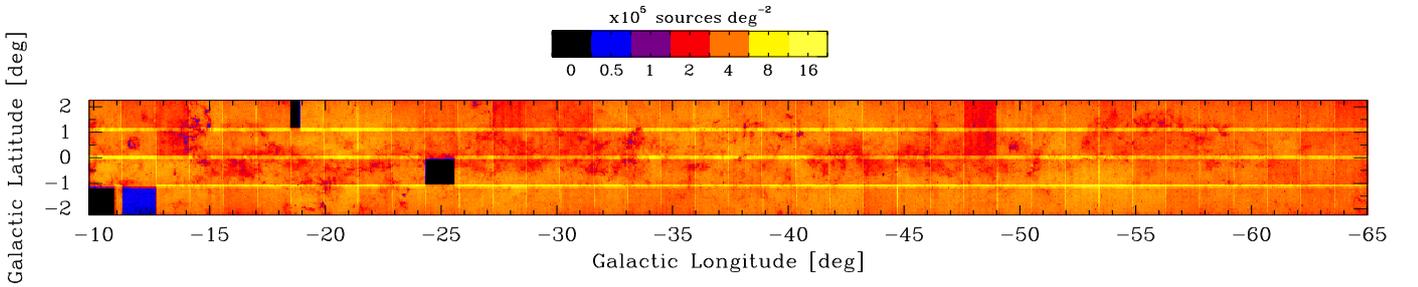}
\caption{Density map  in logarithmic scale showing the  VVV disk area.
  The notation is similar to that presented in Fig.~\ref{fig:bulge}.}
\label{fig:disk}
\end{figure*}

Figures~\ref{fig:bulge}  and  \ref{fig:disk}  show  all  objects  with
stellar flag detected in each tile in the DR1 data. Only stars matched
in the $J$, $H$ and $K_{\rm  s}$ filters have been plotted.  The total
number  of sources found  in the  bulge region  is $7.06~\times~10^7$,
while  the disk  has $9.29~\times~10^7$  sources. The  stellar density
(corrected for the total field size)  is higher in the disk due to the
deeper  photometry in  all five  filters  in comparison  to the  bulge
observations         (see         Table~\ref{tab:strategy},        and
Figs.~\ref{fig:maglim_zy},           \ref{fig:maglim_jh}           and
\ref{fig:maglim_kv}).

Note that these density maps already provide a wealth of information
on the extinction and the structure of the inner Galaxy, which will be
investigated in detail in subsequent papers \citep[e.g.,][Saito et
  al. 2011, in prep.]{2011A&A...534A...3G,2011A&A...534L..14G}.

The regular grid pattern seen in  both the bulge and disk areas is due
to the overlap  of adjacent tiles.  The individual  catalogues used in
Figs.~\ref{fig:bulge}  and \ref{fig:disk}  contain  $\sim10\%$ of  the
total number  of sources twice, contributed by  two independent tiles,
and leading to  the much higher source density  shown in both figures.
The  detection   of  a  significant  percentage  of   sources  on  two
independent  tiles   not  only  helps  to  test   and/or  confirm  the
photometric and astrometric calibration, but also to carry out quality
control.   These density maps  are also  used to  identify problematic
tiles, i.e., incomplete data  readouts, missing tiles and observations
taken under  strongly varying seeing conditions (e.g.,  tiles b216 and
b343).

Moreover, those overlap regions will also be beneficial for one of the
main scientific goals of the VVV Survey. Since the different tiles are
observed independently (with the exception of concatenated tiles), the
variable stars included  on those tiles will obtain  twice as many the
$K_{\rm  s}$  epochs,  and  hence  much better  sampled  light  curves
\citep[see also][]{2011rrls.conf..145C}.

\section{Difference image testing}
\label{sec:dia}

Whilst  DR1 does  not  include photometry  based  on difference  image
analysis  (DIA) we have  tested the  suitability of  the data  for DIA
analysis based on  a modified double-pass version of  the ISIS package
\cite{ala98}  originally developed for  the Angstrom  M31 microlensing
survey \citep{ker10}.

Currently  the  VVV  Science  Verification (SV)  field  comprises  the
largest  number of  epochs and  therefore  most DIA  testing has  been
performed on  this dataset. One  issue with difference imaging  of VVV
data  is that  often  the seeing  is  so good  that  the point  spread
function  (PSF)  is  poorly  sampled  by  the  pixel  size  of  around
$0\farcs34$.   For  good DIA  kernel  convolution  we usually  require
upwards  of  2.5~pixels/FWHM.   In  order to  minimize  undersampling,
instead of convolving a good seeing reference image to a poorer seeing
target we select a poor seeing reference frame and convolve the target
image to it. The DIA image is therefore constructed by minimizing
   \begin{equation}
   D^2_{i,j} = \min \sum_{i,j} [ R_{i,j} - (T \otimes K)_{i,j} + B_{i,j}]^2,
   \end{equation}
where the sum  is over pixel coordinates $(i,j)$, $R$  and $T$ are the
reference and target images, $B$ describes the differential background
between $R$  and $T$,  $K$ is  the convolution kernel  and $D$  is the
resulting difference image (see \cite{ala98} for details of how $K$ is
constructed). Prior to  difference imaging we add the  sky images back
onto the sky-subtracted image stacks.

\begin{figure}
\includegraphics[bb=0cm 0.0cm 25cm 33cm,scale=0.21]{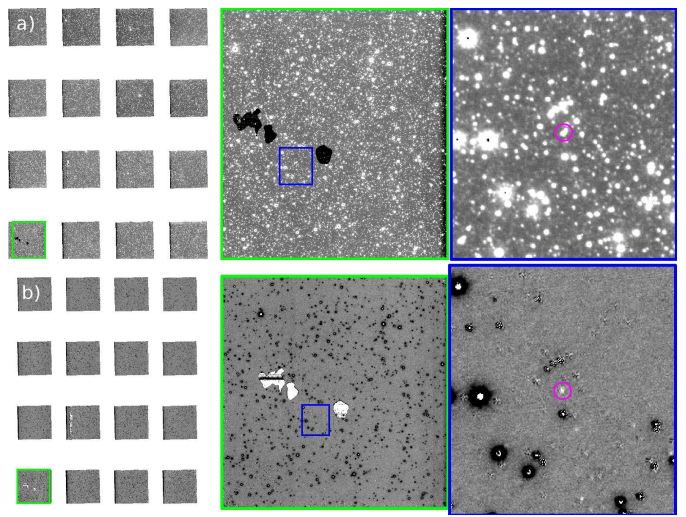}
\caption{(a) A $K_{\rm  s}$ band pawprint from one  VVV SV bulge field
  epoch  showing views  of:  the  full pawprint  (left);  a zoom  into
  Array~1 (middle); and  a further zoom centred on  a circled variable
  object (right).  (b) The bottom row shows  the respective difference
  image views.  Blackened objects  are DIA residuals  around saturated
  stars, which  comprise a non-negligible  fraction of the  image area
  for bulge fields.}
\label{pawdia}
\end{figure}

Figure~\ref{pawdia} shows a $K_S$ band pawprint from one epoch of the
VVV  SV bulge  field. The  lower  row shows  the resulting  difference
images. Whilst the difference  image quality is reasonably clean there
are many black residuals occurring  around the many saturated stars in
the field.  However, variable objects are  easy to pick out  as in the
circled example in the zoomed DIA panel in Fig.~\ref{pawdia}.

\begin{figure}
\includegraphics[bb=.5cm 1.0cm 20cm 28cm,scale=0.205]{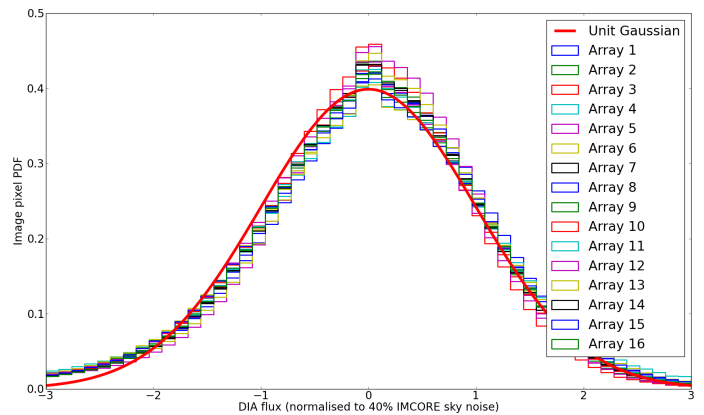}
\caption{Noise histograms for the $K_{\rm s}$ band DIA images shown in
  row  (b) of  Fig.~\ref{pawdia}.  The  $x$-axis is  difference flux
  normalised  to  $40\%$ of  the  sky  noise  level reported  by  {\it
    imcore}. The  smooth function is a unit  Gaussian, which indicates
  that for this image DIA  noise is reasonably Gaussian for all arrays
  and is characterised  by a noise level which is  well below the {\it
    imcore} sky noise estimate.}
\label{diahist}
\end{figure}

The background noise level of  the DIA images is consistent across all
arrays  and performs  well  with  respect to  {\it  imcore} sky  noise
estimates  \citep{2004SPIE.5493..411I}.   In  Fig.~\ref{diahist}  we
plot the pixel  histograms of DIA flux for the  same SV pawprint shown
in  Fig.~\ref{pawdia} for  each of  the  16 arrays.  The fluxes  are
normalised to a  level which is just $40\%$ of  the sky noise estimate
reported  by {\it  imcore}.  In  all cases  the  histograms show  that
difference imaging noise is significantly below that predicted by {\it
  imcore}, as evidenced by the close resemblance of the normalised DIA
flux distributions to a unit Gaussian curve.

The photometric performance  of VVV DIA will be  calibrated in further
detail in a future paper.

\section{Summary}

We  have presented  the VVV  Data  Release 1,  describing the  design,
observations, data processing, data  quality and limitations. The data
are  of very  high  quality,  and represent  a  vast improvement  over
existing  near-IR photometry,  and  are therefore  useful  for a  wide
variety      of      studies      of:      open/embedded      clusters
\citep{2011A&A...532A.131B,2011A&A...531A..73B},   globular   clusters
\citep{2011A&A...527A..81M,2011arXiv1109.1854M},     distance    scale
\citep{2011ApJ...741L..27M},  YSO censuses (Faimali  et al.   2011, in
prep.), brown dwarfs  (Folkes et al.  2011, in  prep.), proper motions
(Minniti  et  al.   2011,  in  prep.), disk  stellar  populations  and
variable stars \citep{2011arXiv1110.3453P},  Galactic structure of the
disk mapping the edge of the stellar disk \citep{2011ApJ...733L..43M},
and  the  inner structure  of  the bulge  \citep{2011A&A...534L..14G},
bulge stellar  populations including metallicity,  extinction and dust
maps        \citep[][Catelan       et        al.         2011       in
  prep]{2011A&A...534A...3G}\footnote{The    Bulge    Extinction   and
  Metallicity  calculator based  on  VVV maps  is  available at:  {\tt
    http://www.eso.org/$\sim$ogonzale/BEAMEC/calculator.php}}
high-energy    sources    \citep[][Masetti    et   al.     2011,    in
  prep.]{greiss2011a,greiss2011b},  background  galaxies (Am\^ores  et
al.  2011,  in prep.), color  transformations between VISTA  and 2MASS
systems  (Soto  et al.   2011,  in prep.),  as  well  as enabling  and
complementing other Galactic structure and stellar population studies,
variable  star  studies  in  clusters  and  in  the  field  (pulsating
variables, eclipsing  binaries and planetary  transits), gravitational
microlensing  studies,  Galactic  Centre studies,  ultra-high-velocity
star  searches, PNe searches,  SN light  echo searches,  QSO searches,
searches  for faint  Solar  System objects  (e.g.,  NEOs, MBAs,  LJ5s,
TNOs), etc.


\begin{acknowledgements}

We  gratefully acknowledge  use of  data  from the  ESO Public  Survey
programme ID 179.B-2002 taken  with the VISTA telescope, data products
from  the Cambridge  Astronomical Survey  Unit, and  funding  from the
FONDAP  Center for Astrophysics  15010003, the  BASAL CATA  Center for
Astrophysics  and Associated  Technologies PFB-06,  the  FONDECYT from
CONICYT, and the Ministry  for the Economy, Development, and Tourism's
Programa  Iniciativa Cient\'{i}fica  Milenio through  grant P07-021-F,
awarded to  The Milky Way  Millennium Nucleus. RKS and  DM acknowledge
financial support  from CONICYT through Gemini  Project No.  32080016.
DM acknowledge support by  Proyecto FONDECYT Regular No.  1090213.  MZ
and OAG acknowledge support  by Proyecto FONDECYT Regular No. 1110393.
JB and  FP are  supported by FONDECYT  Regular No.  1080086.   JRAC is
supported by CONICYT through Gemini Project No. 32090002. ANC received
support  from Comitee  Mixto ESO-Gobierno  de Chile.   MS acknowledges
support  by Proyecto  FONDECYT Regular  No. 3110188  and  Comite Mixto
ESO-Gobierno  de  Chile.   Support  for  RA is  provided  by  Proyecto
FONDECYT  Regular  No.  3100029.   RdG  acknowledges partial  research
support  through  grant 11073001  from  the  National Natural  Science
Foundation  of  China.  MMH  is  supported for  this  work  by the  US
National Science Foundation under  Grant No.  0607497 and 1009550.  GP
acknowledge support  from the Millennium Center  for Supernova Science
through grant  P06-045-F funded by Programa Bicentenario  de Ciencia y
Tecnologƒ¹a de CONICYT.  MC, JAG and ID acknowledge support by Proyecto
FONDECYT Regular No. 1110326. MC is also supported in part by Proyecto
Anillo   ACT-86.    SLF   acknowledges   funding  support   from   the
ESO-Government   of  Chile   Mixed   Committee  2009   and  from   the
Gemini--CONICYT grant  No. 32090014/2009.  BB, BD  and LSJ acknowledge
support  from FAPESP  and  CNPq. EBA  thanks  Funda\c{c}\~{a}o para  a
Ci\^{e}ncia  e Tecnologia (FCT)  under the  grant SFRH/BPD/42239/2007.
RK acknowledges  support from the Centro de  Astrofisica de Valparaiso
and Proyecto DIUV23/2009.


\end{acknowledgements}

\Online

\begin{appendix}
\section{VVV Tile coordinates}
\label{app:tiles}

Here we list the tile  centre coordinates for all VVV pointings. There
are 196  bulge tiles, with names  starting with "b", and  152 tiles in
the  disk area,  having  names starting  with  "d". For  each tile  we
provide   tile   centre  coordinates   in   Equatorial  and   Galactic
coordinates.   All  tiles  have  been  observed  using  the  identical
offsetting strategy,  combining 6 pawprints to  contiguously fill $1.5
\times 1.1$~sq.~deg area. The second last column lists the filters for
which the tile has been completed (i.e., observed within constraints),
and the  last column lists the  number of epochs taken  in $K_{\rm s}$
band within the first observing season (DR1).

\longtab{1}{
\begin{longtable}{lccrrrc}
\caption{\label{tab:tiles} VVV Tile centres.}\\
\hline 
Tile & RA (J2000.0)  & DEC (J2000.0) & longitude & latitude & Filters~~ & $K_{\rm s}$ epochs \\
name & dd:mm:ss.sss  & dd:mm:ss.ss   & degrees~~ & degrees~ & completed & completed         \\
\hline
\endfirsthead
\caption{continued.} \\
\hline
Tile & RA (J2000.0)  & DEC (J2000.0) & longitude & latitude & Filters~~ & $K_{\rm s}$ epochs \\
name & dd:mm:ss.sss  & dd:mm:ss.ss   & degrees~~ & degrees~ & completed & completed         \\
\hline
\endhead
\hline
\endfoot
d001 & 11:43:24.936 & $-$63:31:38.64 & 295.43770 & $-$1.64975 & $ZYJHK_{\rm s}$ & 5 \\
d002 & 11:56:12.576 & $-$63:52:21.00 & 296.89672 & $-$1.64979 & $ZYJHK_{\rm s}$ & 5 \\
d003 & 12:09:17.184 & $-$64:08:46.68 & 298.35572 & $-$1.64971 & $ZYJHK_{\rm s}$ & 5 \\
d004 & 12:22:35.184 & $-$64:20:48.12 & 299.81470 & $-$1.64971 & $ZYJHK_{\rm s}$ & 5 \\
d005 & 12:36:02.640 & $-$64:28:18.84 & 301.27373 & $-$1.64973 & $ZYJHK_{\rm s}$ & 5 \\
d006 & 12:49:35.184 & $-$64:31:14.88 & 302.73271 & $-$1.64977 & $ZYJHK_{\rm s}$ & 5 \\
d007 & 13:03:08.352 & $-$64:29:34.44 & 304.19170 & $-$1.64978 & $ZYJHK_{\rm s}$ & 5 \\
d008 & 13:16:37.632 & $-$64:23:18.24 & 305.65072 & $-$1.64970 & $ZYJHK_{\rm s}$ & 5 \\
d009 & 13:29:58.632 & $-$64:12:30.24 & 307.10972 & $-$1.64971 & ~~~~~$JHK_{\rm s}$ & 5 \\
d010 & 13:43:07.272 & $-$63:57:15.84 & 308.56873 & $-$1.64973 & $ZYJHK_{\rm s}$ & 5 \\
d011 & 13:55:59.856 & $-$63:37:42.60 & 310.02772 & $-$1.64971 & $ZYJHK_{\rm s}$ & 5 \\
d012 & 14:08:33.240 & $-$63:14:00.24 & 311.48673 & $-$1.64970 & $ZYJHK_{\rm s}$ & 5 \\
d013 & 14:20:44.808 & $-$62:46:19.92 & 312.94573 & $-$1.64979 & $ZYJHK_{\rm s}$ & 5 \\
d014 & 14:32:32.496 & $-$62:14:52.80 & 314.40472 & $-$1.64974 & $ZYJHK_{\rm s}$ & 5 \\
d015 & 14:43:42.144 & $-$61:40:33.96 & 315.83598 & $-$1.64972 & $ZYJHK_{\rm s}$ & 4 \\
d016 & 14:54:38.784 & $-$61:02:16.44 & 317.29497 & $-$1.64975 & $ZYJHK_{\rm s}$ & 4 \\
d017 & 15:05:08.712 & $-$60:20:51.00 & 318.75395 & $-$1.64975 & $ZYJHK_{\rm s}$ & 4 \\
d018 & 15:15:11.880 & $-$59:36:30.60 & 320.21293 & $-$1.64975 & $ZYJHK_{\rm s}$ & 4 \\
d019 & 15:24:48.600 & $-$58:49:27.84 & 321.67194 & $-$1.64978 & $ZYJHK_{\rm s}$ & 4 \\
d020 & 15:33:59.400 & $-$57:59:54.60 & 323.13095 & $-$1.64976 & $ZYJHK_{\rm s}$ & 4 \\
d021 & 15:42:45.072 & $-$57:08:02.40 & 324.58996 & $-$1.64971 & $ZYJHK_{\rm s}$ & 4 \\
d022 & 15:51:06.576 & $-$56:14:02.40 & 326.04898 & $-$1.64975 & $ZYJHK_{\rm s}$ & 4 \\
d023 & 15:59:04.920 & $-$55:18:04.32 & 327.50799 & $-$1.64974 & $ZYJHK_{\rm s}$ & 5 \\
d024 & 16:06:41.208 & $-$54:20:17.88 & 328.96694 & $-$1.64974 & $ZYJHK_{\rm s}$ & 5 \\
d025 & 16:13:56.640 & $-$53:20:51.36 & 330.42599 & $-$1.64974 & $ZYJHK_{\rm s}$ & 5 \\
d026 & 16:20:52.320 & $-$52:19:53.40 & 331.88495 & $-$1.64978 & $ZYJHK_{\rm s}$ & 5 \\
d027 & 16:27:29.400 & $-$51:17:30.84 & 333.34393 & $-$1.64976 & ~~~~~$JHK_{\rm s}$ &  2 \\
d028 & 16:33:49.032 & $-$50:13:50.52 & 334.80299 & $-$1.64976 & ~~~~~$JHK_{\rm s}$ &  2\\
d029 & 16:39:52.248 & $-$49:08:58.92 & 336.26199 & $-$1.64971 & ~~~~~$JHK_{\rm s}$ &  2\\
d030 & 16:45:40.128 & $-$48:03:01.80 & 337.72099 & $-$1.64973 & $ZYJHK_{\rm s}$ & 2 \\
d031 & 16:51:13.632 & $-$46:56:04.20 & 339.17999 & $-$1.64971 & $ZYJHK_{\rm s}$ & 2 \\
d032 & 16:56:33.720 & $-$45:48:11.16 & 340.63896 & $-$1.64975 & $ZYJHK_{\rm s}$ & 2 \\
d033 & 17:01:41.256 & $-$44:39:26.64 & 342.09795 & $-$1.64972 & $ZYJHK_{\rm s}$ & 1 \\
d034 & 17:06:37.104 & $-$43:29:54.96 & 343.55695 & $-$1.64975 & $ZYJHK_{\rm s}$ & 1 \\
d035 & 17:11:22.032 & $-$42:19:39.72 & 345.01595 & $-$1.64979 & $ZYJHK_{\rm s}$ & 1 \\
d036 & 17:15:56.760 & $-$41:08:44.16 & 346.47495 & $-$1.64979 & ~~~~~$JHK_{\rm s}$ & 1 \\
d037 & 17:20:21.984 & $-$39:57:11.16 & 347.93400 & $-$1.64973 & ~~~~~$JHK_{\rm s}$ & 1 \\
d038 & 17:24:38.352 & $-$38:45:04.32 & 349.39294 & $-$1.64975 & ~~~~~$JHK_{\rm s}$ & 1 \\
d039 & 11:45:52.488 & $-$62:28:17.40 & 295.43747 & $-$0.55759 & $ZYJHK_{\rm s}$ & 5 \\
d040 & 11:58:14.160 & $-$62:48:15.12 & 296.89617 & $-$0.55758 & $ZYJHK_{\rm s}$ & 5 \\
d041 & 12:10:50.928 & $-$63:04:04.80 & 298.35479 & $-$0.55753 & $ZYJHK_{\rm s}$ & 5 \\
d042 & 12:23:39.672 & $-$63:15:39.60 & 299.81350 & $-$0.55756 & $ZYJHK_{\rm s}$ & 5 \\
d043 & 12:36:36.744 & $-$63:22:53.40 & 301.27213 & $-$0.55754 & $ZYJHK_{\rm s}$ & 5 \\
d044 & 12:49:38.376 & $-$63:25:42.96 & 302.73081 & $-$0.55755 & $ZYJHK_{\rm s}$ & 5 \\
d045 & 13:02:40.560 & $-$63:24:06.84 & 304.18948 & $-$0.55760 & $ZYJHK_{\rm s}$ & 5 \\
d046 & 13:15:39.288 & $-$63:18:05.40 & 305.64814 & $-$0.55754 & $ZYJHK_{\rm s}$ & 5 \\
d047 & 13:28:30.696 & $-$63:07:42.24 & 307.10682 & $-$0.55756 & ~~~~~$JHK_{\rm s}$ &   5 \\
d048 & 13:41:11.112 & $-$62:53:02.04 & 308.56549 & $-$0.55756 & ~~~~~$JHK_{\rm s}$ &   5\\
d049 & 13:53:37.224 & $-$62:34:12.00 & 310.02413 & $-$0.55760 & $ZYJHK_{\rm s}$ & 5 \\
d050 & 14:05:46.176 & $-$62:11:20.04 & 311.48283 & $-$0.55756 & $ZYJHK_{\rm s}$ & 5 \\
d051 & 14:17:35.520 & $-$61:44:36.60 & 312.94152 & $-$0.55757 & $ZYJHK_{\rm s}$ & 5 \\
d052 & 14:29:03.312 & $-$61:14:12.48 & 314.40014 & $-$0.55761 & $ZYJHK_{\rm s}$ & 5 \\
d053 & 14:40:08.135 & $-$60:40:18.48 & 315.85883 & $-$0.55752 & $ZYJHK_{\rm s}$ & 4 \\
d054 & 14:50:49.032 & $-$60:03:07.56 & 317.31750 & $-$0.55758 & $ZYJHK_{\rm s}$ & 4 \\
d055 & 15:01:05.424 & $-$59:22:51.24 & 318.77616 & $-$0.55761 & $ZYJHK_{\rm s}$ & 4 \\
d056 & 15:10:57.120 & $-$58:39:41.40 & 320.23482 & $-$0.55759 & $ZYJHK_{\rm s}$ & 3 \\
d057 & 15:20:24.264 & $-$57:53:49.92 & 321.69349 & $-$0.55755 & $ZYJHK_{\rm s}$ & 4 \\
d058 & 15:29:27.264 & $-$57:05:28.32 & 323.15217 & $-$0.55754 & $ZYJHK_{\rm s}$ & 4 \\
d059 & 15:38:06.720 & $-$56:14:47.40 & 324.61087 & $-$0.55755 & $ZYJHK_{\rm s}$ & 4 \\
d060 & 15:46:23.352 & $-$55:21:57.60 & 326.06950 & $-$0.55754 & $ZYJHK_{\rm s}$ & 4 \\
d061 & 15:54:18.096 & $-$54:27:08.64 & 327.52817 & $-$0.55761 & $ZYJHK_{\rm s}$ & 5 \\
d062 & 16:01:51.840 & $-$53:30:29.16 & 328.98684 & $-$0.55756 & $ZYJHK_{\rm s}$ & 5 \\
d063 & 16:09:05.616 & $-$52:32:08.16 & 330.44549 & $-$0.55755 & $ZYJHK_{\rm s}$ & 5 \\
d064 & 16:16:00.456 & $-$51:32:13.20 & 331.90419 & $-$0.55754 & $ZYJHK_{\rm s}$ & 5 \\
d065 & 16:22:37.368 & $-$50:30:51.84 & 333.36286 & $-$0.55756 & ~~~~~$JHK_{\rm s}$ & 2 \\
d066 & 16:28:57.360 & $-$49:28:10.56 & 334.82153 & $-$0.55755 & ~~~~~$JHK_{\rm s}$ & 2 \\
d067 & 16:35:01.416 & $-$48:24:15.84 & 336.28015 & $-$0.55756 & ~~~~~$JHK_{\rm s}$ & 2 \\
d068 & 16:40:50.520 & $-$47:19:13.08 & 337.73882 & $-$0.55761 & $ZYJHK_{\rm s}$ & 2 \\
d069 & 16:46:25.560 & $-$46:13:07.32 & 339.19753 & $-$0.55758 & $ZYJHK_{\rm s}$ & 2 \\
d070 & 16:51:47.400 & $-$45:06:03.96 & 340.65613 & $-$0.55757 & $ZYJHK_{\rm s}$ & 2 \\
d071 & 16:56:56.928 & $-$43:58:06.96 & 342.11483 & $-$0.55761 & $ZYJHK_{\rm s}$ & 1 \\
d072 & 17:01:54.864 & $-$42:49:20.28 & 343.57351 & $-$0.55754 & $ZYJHK_{\rm s}$ & 1 \\
d073 & 17:06:42.000 & $-$41:39:48.24 & 345.03215 & $-$0.55756 & $ZYJHK_{\rm s}$ & 1 \\
d074 & 17:11:19.032 & $-$40:29:33.72 & 346.49083 & $-$0.55759 & ~~~~~$JHK_{\rm s}$ & 1 \\
d075 & 17:15:46.608 & $-$39:18:39.96 & 347.94953 & $-$0.55759 & ~~~~~$JHK_{\rm s}$ & 1 \\
d076 & 17:20:05.328 & $-$38:07:09.84 & 349.40822 & $-$0.55752 & ~~~~~$JHK_{\rm s}$ & 1 \\
d077 & 11:48:10.080 & $-$61:24:47.16 & 295.43749 & 0.53461 & $ZYJHK_{\rm s}$ & 5 \\
d078 & 12:00:07.584 & $-$61:44:03.84 & 296.89636 & 0.53458 & $ZYJHK_{\rm s}$ & 5 \\
d079 & 12:12:18.648 & $-$61:59:20.40 & 298.35521 & 0.53456 & $ZYJHK_{\rm s}$ & 5 \\
d080 & 12:24:40.392 & $-$62:10:30.00 & 299.81408 & 0.53461 & $ZYJHK_{\rm s}$ & 5 \\
d081 & 12:37:09.600 & $-$62:17:27.96 & 301.27295 & 0.53465 & $ZYJHK_{\rm s}$ & 5 \\
d082 & 12:49:42.840 & $-$62:20:11.40 & 302.73182 & 0.53458 & $ZYJHK_{\rm s}$ & 5 \\
d083 & 13:02:16.560 & $-$62:18:38.16 & 304.19066 & 0.53466 & $ZYJHK_{\rm s}$ & 5 \\
d084 & 13:14:47.232 & $-$62:12:50.04 & 305.64955 & 0.53458 & $ZYJHK_{\rm s}$ & 5 \\
d085 & 13:27:11.328 & $-$62:02:48.84 & 307.10837 & 0.53460 & $ZYJHK_{\rm s}$ & 4 \\
d086 & 13:39:25.632 & $-$61:48:39.24 & 308.56725 & 0.53465 & $ZYJHK_{\rm s}$ & 4 \\
d087 & 13:51:27.144 & $-$61:30:28.08 & 310.02613 & 0.53456 & $ZYJHK_{\rm s}$ & 5 \\
d088 & 14:03:13.176 & $-$61:08:22.20 & 311.48497 & 0.53458 & $ZYJHK_{\rm s}$ & 5 \\
d089 & 14:14:41.544 & $-$60:42:30.60 & 312.94386 & 0.53465 & $ZYJHK_{\rm s}$ & 4 \\
d090 & 14:25:50.400 & $-$60:13:03.72 & 314.40272 & 0.53463 & $ZYJHK_{\rm s}$ & 4 \\
d091 & 14:36:38.352 & $-$59:40:11.64 & 315.86159 & 0.53462 & $ZYJHK_{\rm s}$ & 4 \\
d092 & 14:47:04.392 & $-$59:04:05.52 & 317.32043 & 0.53461 & $ZYJHK_{\rm s}$ & 4 \\
d093 & 14:57:07.920 & $-$58:24:56.16 & 318.77932 & 0.53465 & $ZYJHK_{\rm s}$ & 3 \\
d094 & 15:06:48.648 & $-$57:42:55.44 & 320.23818 & 0.53459 & $ZYJHK_{\rm s}$ & 3 \\
d095 & 15:16:06.552 & $-$56:58:13.80 & 321.69700 & 0.53459 & $ZYJHK_{\rm s}$ & 5 \\
d096 & 15:25:01.920 & $-$56:11:02.04 & 323.15588 & 0.53461 & $ZYJHK_{\rm s}$ & 5 \\
d097 & 15:33:35.208 & $-$55:21:30.60 & 324.61479 & 0.53459 & $ZYJHK_{\rm s}$ & 4 \\
d098 & 15:41:46.968 & $-$54:29:49.20 & 326.07365 & 0.53465 & $ZYJHK_{\rm s}$ & 4 \\
d099 & 15:49:37.968 & $-$53:36:07.56 & 327.53250 & 0.53463 & $ZYJHK_{\rm s}$ & 4 \\
d100 & 15:57:09.024 & $-$52:40:34.32 & 328.99135 & 0.53458 & $ZYJHK_{\rm s}$ & 4 \\
d101 & 16:04:20.976 & $-$51:43:17.40 & 330.45023 & 0.53459 & $ZYJHK_{\rm s}$ & 5 \\
d102 & 16:11:14.736 & $-$50:44:24.72 & 331.90911 & 0.53460 & $ZYJHK_{\rm s}$ & 5 \\
d103 & 16:17:51.216 & $-$49:44:03.48 & 333.36798 & 0.53460 & $ZYJHK_{\rm s}$ & 3 \\
d104 & 16:24:11.328 & $-$48:42:20.16 & 334.82687 & 0.53461 & $ZYJHK_{\rm s}$ & 3 \\
d105 & 16:30:15.960 & $-$47:39:21.24 & 336.28568 & 0.53460 & $ZYJHK_{\rm s}$ & 3 \\
d106 & 16:36:06.000 & $-$46:35:12.12 & 337.74450 & 0.53460 & $ZYJHK_{\rm s}$ & 3 \\
d107 & 16:41:42.312 & $-$45:29:57.84 & 339.20338 & 0.53460 & ~~~~~$JHK_{\rm s}$ & 1 \\
d108 & 16:47:05.712 & $-$44:23:43.44 & 340.66229 & 0.53457 & ~~~~~$JHK_{\rm s}$ & 1 \\
d109 & 16:52:16.944 & $-$43:16:33.24 & 342.12118 & 0.53461 & ~~~~~$JHK_{\rm s}$ & 1 \\
d110 & 16:57:16.776 & $-$42:08:31.56 & 343.58005 & 0.53462 & $ZYJHK_{\rm s}$ & 1 \\
d111 & 17:02:05.904 & $-$40:59:42.36 & 345.03883 & 0.53459 & $ZYJHK_{\rm s}$ & 1 \\
d112 & 17:06:45.024 & $-$39:50:08.52 & 346.49771 & 0.53457 & $ZYJHK_{\rm s}$ & 1 \\
d113 & 17:11:14.736 & $-$38:39:53.28 & 347.95664 & 0.53461 & $ZYJHK_{\rm s}$ & 2 \\
d114 & 17:15:35.640 & $-$37:29:00.24 & 349.41546 & 0.53460 & $ZYJHK_{\rm s}$ & 2 \\
d115 & 11:50:18.720 & $-$60:21:09.00 & 295.43768 & 1.62680 & $ZYJHK_{\rm s}$ & 5 \\
d116 & 12:01:53.760 & $-$60:39:47.52 & 296.89732 & 1.62677 & $ZYJHK_{\rm s}$ & 5 \\
d117 & 12:13:40.992 & $-$60:54:32.75 & 298.35689 & 1.62684 & $ZYJHK_{\rm s}$ & 5 \\
d118 & 12:25:37.800 & $-$61:05:19.68 & 299.81648 & 1.62674 & $ZYJHK_{\rm s}$ & 5 \\
d119 & 12:37:41.304 & $-$61:12:02.52 & 301.27608 & 1.62684 & $ZYJHK_{\rm s}$ & 5 \\
d120 & 12:49:48.408 & $-$61:14:39.48 & 302.73567 & 1.62683 & $ZYJHK_{\rm s}$ & 5 \\
d121 & 13:01:55.944 & $-$61:13:09.12 & 304.19526 & 1.62680 & $ZYJHK_{\rm s}$ & 5 \\
d122 & 13:14:00.720 & $-$61:07:32.15 & 305.65484 & 1.62675 & $ZYJHK_{\rm s}$ & 5 \\
d123 & 13:25:59.640 & $-$60:57:50.40 & 307.11447 & 1.62681 & $ZYJHK_{\rm s}$ & 4 \\
d124 & 13:37:49.704 & $-$60:44:08.88 & 308.57402 & 1.62675 & $ZYJHK_{\rm s}$ & 4 \\
d125 & 13:49:28.248 & $-$60:26:32.28 & 310.03361 & 1.62680 & $ZYJHK_{\rm s}$ & 5 \\
d126 & 14:00:52.872 & $-$60:05:08.16 & 311.49324 & 1.62678 & $ZYJHK_{\rm s}$ & 5 \\
d127 & 14:12:01.440 & $-$59:40:04.44 & 312.95280 & 1.62677 & $ZYJHK_{\rm s}$ & 4 \\
d128 & 14:22:52.272 & $-$59:11:29.76 & 314.41240 & 1.62684 & $ZYJHK_{\rm s}$ & 4 \\
d129 & 14:33:24.024 & $-$58:39:34.56 & 315.87197 & 1.62678 & $ZYJHK_{\rm s}$ & 4 \\
d130 & 14:43:35.688 & $-$58:04:28.20 & 317.33156 & 1.62679 & $ZYJHK_{\rm s}$ & 4 \\
d131 & 14:53:26.616 & $-$57:26:21.48 & 318.79117 & 1.62676 & $ZYJHK_{\rm s}$ & 3 \\
d132 & 15:02:56.424 & $-$56:45:24.48 & 320.25077 & 1.62679 & $ZYJHK_{\rm s}$ & 3 \\
d133 & 15:12:05.040 & $-$56:01:48.00 & 321.71037 & 1.62674 & $ZYJHK_{\rm s}$ & 5 \\
d134 & 15:20:52.560 & $-$55:15:41.76 & 323.16993 & 1.62676 & $ZYJHK_{\rm s}$ & 5 \\
d135 & 15:29:19.344 & $-$54:27:15.48 & 324.62955 & 1.62682 & $ZYJHK_{\rm s}$ & 4 \\
d136 & 15:37:25.872 & $-$53:36:39.24 & 326.08912 & 1.62676 & $ZYJHK_{\rm s}$ & 4 \\
d137 & 15:45:12.720 & $-$52:44:01.32 & 327.54872 & 1.62676 & $ZYJHK_{\rm s}$ & 4 \\
d138 & 15:52:40.584 & $-$51:49:30.36 & 329.00835 & 1.62678 & $ZYJHK_{\rm s}$ & 4 \\
d139 & 15:59:50.184 & $-$50:53:14.64 & 330.46790 & 1.62680 & $ZYJHK_{\rm s}$ & 5 \\
d140 & 16:06:42.360 & $-$49:55:21.36 & 331.92752 & 1.62681 & $ZYJHK_{\rm s}$ & 5 \\
d141 & 16:13:17.904 & $-$48:55:57.72 & 333.38714 & 1.62678 & $ZYJHK_{\rm s}$ & 3 \\
d142 & 16:19:37.608 & $-$47:55:10.20 & 334.84670 & 1.62678 & $ZYJHK_{\rm s}$ & 3 \\
d143 & 16:25:42.312 & $-$46:53:04.92 & 336.30624 & 1.62678 & $ZYJHK_{\rm s}$ & 3 \\
d144 & 16:31:32.832 & $-$45:49:46.92 & 337.76590 & 1.62684 & $ZYJHK_{\rm s}$ & 3 \\
d145 & 16:37:09.936 & $-$44:45:22.32 & 339.22547 & 1.62682 & $ZYJHK_{\rm s}$ & 1 \\
d146 & 16:42:34.392 & $-$43:39:55.44 & 340.68506 & 1.62681 & ~~~~~$JHK_{\rm s}$ & 1 \\
d147 & 16:47:46.920 & $-$42:33:30.96 & 342.14462 & 1.62677 & ~~~~~$JHK_{\rm s}$ & 1 \\
d148 & 16:52:48.216 & $-$41:26:12.48 & 343.60424 & 1.62683 & ~~~~~$JHK_{\rm s}$ & 1 \\
d149 & 16:57:38.976 & $-$40:18:04.32 & 345.06387 & 1.62680 & $ZYJHK_{\rm s}$ & 1 \\
d150 & 17:02:19.800 & $-$39:09:10.08 & 346.52343 & 1.62677 & $ZYJHK_{\rm s}$ & 1 \\
d151 & 17:06:51.312 & $-$37:59:32.64 & 347.98303 & 1.62675 & $ZYJHK_{\rm s}$ & 2 \\
d152 & 17:11:14.064 & $-$36:49:15.24 & 349.44260 & 1.62674 & $ZYJHK_{\rm s}$ & 2 \\
b201 & 18:04:24.384 & $-$41:44:53.52 & 350.74816 & $-$9.68974 & $ZYJHK_{\rm s}$ & 1 \\
b202 & 18:08:00.144 & $-$40:27:29.88 & 352.22619 & $-$9.68971 & $ZYJHK_{\rm s}$ & 1 \\
b203 & 18:11:29.496 & $-$39:09:52.92 & 353.70409 & $-$9.68973 & $ZYJHK_{\rm s}$ & 1 \\
b204 & 18:14:52.992 & $-$37:52:03.36 & 355.18207 & $-$9.68974 & $ZYJHK_{\rm s}$ & 1 \\
b205 & 18:18:11.136 & $-$36:34:02.64 & 356.66012 & $-$9.68976 & $ZYJHK_{\rm s}$ & 1 \\
b206 & 18:21:24.360 & $-$35:15:52.20 & 358.13813 & $-$9.68975 & $ZYJHK_{\rm s}$ & 1 \\
b207 & 18:24:33.096 & $-$33:57:33.48 & 359.61607 & $-$9.68977 & $ZYJHK_{\rm s}$ & 2 \\
b208 & 18:27:37.728 & $-$32:39:07.20 & 1.09399 & $-$9.68974 & $ZYJHK_{\rm s}$ &   2 \\
b209 & 18:30:38.640 & $-$31:20:34.08 & 2.57200 & $-$9.68971 & $ZYJHK_{\rm s}$ &   2 \\
b210 & 18:33:36.168 & $-$30:01:55.56 & 4.04998 & $-$9.68973 & $ZYJHK_{\rm s}$ & 2 \\
b211 & 18:36:30.624 & $-$28:43:12.36 & 5.52796 & $-$9.68978 & $ZYJHK_{\rm s}$ & 2 \\
b212 & 18:39:22.272 & $-$27:24:25.20 & 7.00593 & $-$9.68975 & $ZYJHK_{\rm s}$ & 3 \\
b213 & 18:42:11.424 & $-$26:05:34.80 & 8.48396 & $-$9.68974 & $ZYJHK_{\rm s}$ & 2 \\
b214 & 18:44:58.320 & $-$24:46:42.24 & 9.96193 & $-$9.68974 & $ZYJHK_{\rm s}$ & 3 \\
b215 & 17:59:15.960 & $-$41:13:55.92 & 350.74595 & $-$8.59756 & $ZYJHK_{\rm s}$ & 1 \\
b216 & 18:02:55.992 & $-$39:57:07.92 & 352.21956 & $-$8.59753 & $ZYJHK_{\rm s}$ & 1 \\
b217 & 18:06:29.472 & $-$38:40:04.08 & 353.69327 & $-$8.59756 & $ZYJHK_{\rm s}$ & 1 \\
b218 & 18:09:56.880 & $-$37:22:46.56 & 355.16684 & $-$8.59757 & $ZYJHK_{\rm s}$ & 1 \\
b219 & 18:13:18.768 & $-$36:05:16.07 & 356.64051 & $-$8.59760 & $ZYJHK_{\rm s}$ & 1 \\
b220 & 18:16:35.568 & $-$34:47:34.08 & 358.11423 & $-$8.59759 & $ZYJHK_{\rm s}$ & 1 \\
b221 & 18:19:47.688 & $-$33:29:42.36 & 359.58781 & $-$8.59757 & $ZYJHK_{\rm s}$ & 2 \\
b222 & 18:22:55.560 & $-$32:11:41.28 & 1.06151 & $-$8.59755 & $ZYJHK_{\rm s}$ &   2 \\
b223 & 18:25:59.544 & $-$30:53:32.28 & 2.53522 & $-$8.59757 & $ZYJHK_{\rm s}$ &   2 \\
b224 & 18:28:59.952 & $-$29:35:16.80 & 4.00880 & $-$8.59759 & $ZYJHK_{\rm s}$ & 2 \\
b225 & 18:31:57.120 & $-$28:16:54.84 & 5.48250 & $-$8.59755 & $ZYJHK_{\rm s}$ & 2 \\
b226 & 18:34:51.360 & $-$26:58:27.84 & 6.95620 & $-$8.59757 & $ZYJHK_{\rm s}$ & 3 \\
b227 & 18:37:42.912 & $-$25:39:56.88 & 8.42977 & $-$8.59756 & $ZYJHK_{\rm s}$ & 2 \\
b228 & 18:40:32.088 & $-$24:21:21.96 & 9.90350 & $-$8.59757 & $ZYJHK_{\rm s}$ & 3 \\
b229 & 17:54:12.456 & $-$40:42:07.56 & 350.74383 & $-$7.50542 & \,\,\,\,$YJHK_{\rm s}$ & 1 \\
b230 & 17:57:56.496 & $-$39:25:54.48 & 352.21380 & $-$7.50537 & ~~~~~$JHK_{\rm s}$ & -- \\
b231 & 18:01:33.792 & $-$38:09:24.48 & 353.68363 & $-$7.50537 & ~~~~~$JHK_{\rm s}$ & 1 \\
b232 & 18:05:04.920 & $-$36:52:38.28 & 355.15359 & $-$7.50541 & ~~~~~$JHK_{\rm s}$ & 1 \\
b233 & 18:08:30.312 & $-$35:35:38.04 & 356.62342 & $-$7.50539 & ~~~~~$JHK_{\rm s}$ & 1 \\
b234 & 18:11:50.472 & $-$34:18:24.48 & 358.09337 & $-$7.50535 & ~~~~~$JHK_{\rm s}$ & 1 \\
b235 & 18:15:05.832 & $-$33:00:59.76 & 359.56322 & $-$7.50542 & ~~~~~$JHK_{\rm s}$ & 2 \\
b236 & 18:18:16.752 & $-$31:43:24.24 & 1.03312 & $-$7.50538 & $ZYJHK_{\rm s}$ & 2 \\
b237 & 18:21:23.640 & $-$30:25:39.36 & 2.50307 & $-$7.50541 & $ZYJHK_{\rm s}$ & 2 \\
b238 & 18:24:26.808 & $-$29:07:46.20 & 3.97300 & $-$7.50540 & ~~~~~$JHK_{\rm s}$ & 2 \\
b239 & 18:27:26.568 & $-$27:49:45.84 & 5.44287 & $-$7.50536 & ~~~~~$JHK_{\rm s}$ & 2 \\
b240 & 18:30:23.256 & $-$26:31:39.36 & 6.91271 & $-$7.50540 & ~~~~~$JHK_{\rm s}$ & 2 \\
b241 & 18:33:17.136 & $-$25:13:27.12 & 8.38261 & $-$7.50541 & ~~~~~$JHK_{\rm s}$ & 3 \\
b242 & 18:36:08.472 & $-$23:55:10.20 & 9.85251 & $-$7.50542 & ~~~~~$JHK_{\rm s}$ & 2 \\
b243 & 17:49:13.848 & $-$40:09:29.16 & 350.74206 & $-$6.41324 & \,\,\,\,$YJHK_{\rm s}$ & 1 \\
b244 & 17:53:01.608 & $-$38:53:51.72 & 352.20875 & $-$6.41323 & ~~~~~$JHK_{\rm s}$ & -- \\
b245 & 17:56:42.504 & $-$37:37:54.84 & 353.67546 & $-$6.41323 & ~~~~~$JHK_{\rm s}$ & 1 \\
b246 & 18:00:17.064 & $-$36:21:40.32 & 355.14219 & $-$6.41321 & ~~~~~$JHK_{\rm s}$ & 1 \\
b247 & 18:03:45.792 & $-$35:05:09.96 & 356.60888 & $-$6.41323 & ~~~~~$JHK_{\rm s}$ & 1 \\
b248 & 18:07:09.120 & $-$33:48:25.20 & 358.07550 & $-$6.41322 & ~~~~~$JHK_{\rm s}$ & 1 \\
b249 & 18:10:27.504 & $-$32:31:27.12 & 359.54218 & $-$6.41323 & ~~~~~$JHK_{\rm s}$ & 2 \\
b250 & 18:13:41.328 & $-$31:14:17.16 & 1.00886 & $-$6.41325 & $ZYJHK_{\rm s}$ & 2 \\
b251 & 18:16:50.952 & $-$29:56:56.04 & 2.47562 & $-$6.41319 & ~~~~~$JHK_{\rm s}$ &   2 \\
b252 & 18:19:56.736 & $-$28:39:25.92 & 3.94224 & $-$6.41326 & ~~~~~$JHK_{\rm s}$ &   2 \\
b253 & 18:22:58.968 & $-$27:21:46.80 & 5.40892 & $-$6.41319 & ~~~~~$JHK_{\rm s}$ & 2 \\
b254 & 18:25:58.008 & $-$26:04:00.12 & 6.87563 & $-$6.41325 & ~~~~~$JHK_{\rm s}$ & 2 \\
b255 & 18:28:54.072 & $-$24:46:06.60 & 8.34231 & $-$6.41319 & ~~~~~$JHK_{\rm s}$ & 3 \\
b256 & 18:31:47.496 & $-$23:28:07.32 & 9.80903 & $-$6.41325 & ~~~~~$JHK_{\rm s}$ & 2 \\
b257 & 17:44:20.112 & $-$39:36:02.16 & 350.74076 & $-$5.32104 & ~~~~~$JHK_{\rm s}$ & 1 \\
b258 & 17:48:11.328 & $-$38:20:59.64 & 352.20485 & $-$5.32102 & ~~~~~$JHK_{\rm s}$ & 1 \\
b259 & 17:51:55.560 & $-$37:05:36.24 & 353.66885 & $-$5.32101 & ~~~~~$JHK_{\rm s}$ & -- \\
b260 & 17:55:33.360 & $-$35:49:53.40 & 355.13291 & $-$5.32104 & ~~~~~$JHK_{\rm s}$ & -- \\
b261 & 17:59:05.184 & $-$34:33:52.92 & 356.59692 & $-$5.32103 & ~~~~~$JHK_{\rm s}$ & -- \\
b262 & 18:02:31.512 & $-$33:17:36.24 & 358.06096 & $-$5.32105 & ~~~~~$JHK_{\rm s}$ & -- \\
b263 & 18:05:52.752 & $-$32:01:04.80 & 359.52500 & $-$5.32104 & ~~~~~$JHK_{\rm s}$ & -- \\
b264 & 18:09:09.288 & $-$30:44:20.04 & 0.98899 & $-$5.32099 & ~~~~~$JHK_{\rm s}$ & -- \\
b265 & 18:12:21.528 & $-$29:27:23.40 & 2.45295 & $-$5.32106 & ~~~~~$JHK_{\rm s}$ & -- \\
b266 & 18:15:29.784 & $-$28:10:15.24 & 3.91703 & $-$5.32105 & ~~~~~$JHK_{\rm s}$ & -- \\
b267 & 18:18:34.368 & $-$26:52:57.36 & 5.38103 & $-$5.32101 & ~~~~~$JHK_{\rm s}$ & -- \\
b268 & 18:21:35.616 & $-$25:35:30.48 & 6.84507 & $-$5.32101 & ~~~~~$JHK_{\rm s}$ & -- \\
b269 & 18:24:33.792 & $-$24:17:55.68 & 8.30909 & $-$5.32100 & ~~~~~$JHK_{\rm s}$ & 1 \\
b270 & 18:27:29.184 & $-$23:00:14.04 & 9.77309 & $-$5.32107 & $ZYJHK_{\rm s}$ & 1 \\
b271 & 17:39:31.128 & $-$39:01:49.44 & 350.73953 & $-$4.22883 & ~~~~~$JHK_{\rm s}$ & 1 \\
b272 & 17:43:25.536 & $-$37:47:22.20 & 352.20141 & $-$4.22884 & ~~~~~$JHK_{\rm s}$ & 1 \\
b273 & 17:47:12.888 & $-$36:32:31.92 & 353.66332 & $-$4.22886 & ~~~~~$JHK_{\rm s}$ & 1 \\
b274 & 17:50:53.688 & $-$35:17:20.76 & 355.12516 & $-$4.22890 & ~~~~~$JHK_{\rm s}$ & 1 \\
b275 & 17:54:28.416 & $-$34:01:49.80 & 356.58709 & $-$4.22886 & ~~~~~$JHK_{\rm s}$ & 1 \\
b276 & 17:57:57.528 & $-$32:46:01.20 & 358.04898 & $-$4.22882 & ~~~~~$JHK_{\rm s}$ & 1 \\
b277 & 18:01:21.456 & $-$31:29:56.40 & 359.51088 & $-$4.22881 & ~~~~~$JHK_{\rm s}$ & 1 \\
b278 & 18:04:40.584 & $-$30:13:36.84 & 0.97275 & $-$4.22884 & ~~~~~$JHK_{\rm s}$ & 1 \\
b279 & 18:07:55.272 & $-$28:57:03.60 & 2.43463 & $-$4.22884 & ~~~~~$JHK_{\rm s}$ & 1 \\
b280 & 18:11:05.880 & $-$27:40:17.76 & 3.89659 & $-$4.22886 & ~~~~~$JHK_{\rm s}$ & 1 \\
b281 & 18:14:12.696 & $-$26:23:20.76 & 5.35849 & $-$4.22883 & ~~~~~$JHK_{\rm s}$ & 1 \\
b282 & 18:17:16.056 & $-$25:06:13.68 & 6.82039 & $-$4.22886 & ~~~~~$JHK_{\rm s}$ & 1 \\
b283 & 18:20:16.224 & $-$23:48:57.60 & 8.28222 & $-$4.22888 & ~~~~~$JHK_{\rm s}$ & -- \\
b284 & 18:23:13.488 & $-$22:31:32.88 & 9.74416 & $-$4.22889 & ~~~~~$JHK_{\rm s}$ & -- \\
b285 & 17:34:46.896 & $-$38:26:51.72 & 350.73871 & $-$3.13666 & ~~~~~$JHK_{\rm s}$ & -- \\
b286 & 17:38:44.256 & $-$37:12:59.40 & 352.19896 & $-$3.13670 & ~~~~~$JHK_{\rm s}$ & -- \\
b287 & 17:42:34.488 & $-$35:58:41.88 & 353.65931 & $-$3.13670 & ~~~~~$JHK_{\rm s}$ & -- \\
b288 & 17:46:18.096 & $-$34:44:01.68 & 355.11962 & $-$3.13673 & $ZYJHK_{\rm s}$ & -- \\
b289 & 17:49:55.536 & $-$33:29:00.24 & 356.57994 & $-$3.13668 & $ZYJHK_{\rm s}$ & 1 \\
b290 & 17:53:27.288 & $-$32:13:39.72 & 358.04023 & $-$3.13673 & $ZYJHK_{\rm s}$ & 1 \\
b291 & 17:56:53.736 & $-$30:58:01.20 & 359.50054 & $-$3.13672 & $ZYJHK_{\rm s}$ & -- \\
b292 & 18:00:15.264 & $-$29:42:06.12 & 0.96088 & $-$3.13663 & $ZY$ & -- \\
b293 & 18:03:32.280 & $-$28:25:56.28 & 2.42120 & $-$3.13666 & $ZY$ & -- \\
b294 & 18:06:45.096 & $-$27:09:32.75 & 3.88150 & $-$3.13671 & $ZYJHK_{\rm s}$ & -- \\
b295 & 18:09:54.024 & $-$25:52:56.64 & 5.34179 & $-$3.13672 & $ZYJHK_{\rm s}$ & -- \\
b296 & 18:12:59.352 & $-$24:36:09.00 & 6.80204 & $-$3.13666 & $ZYJHK_{\rm s}$ & -- \\
b297 & 18:16:01.416 & $-$23:19:10.92 & 8.26235 & $-$3.13668 & $ZYJHK_{\rm s}$ & 1 \\
b298 & 18:19:00.456 & $-$22:02:03.12 & 9.72271 & $-$3.13666 & ~~~~~$JHK_{\rm s}$ & 1 \\
b299 & 17:30:07.272 & $-$37:51:11.88 & 350.73789 & $-$2.04453 & ~~~~~$JHK_{\rm s}$ & -- \\
b300 & 17:34:07.344 & $-$36:37:53.76 & 352.19711 & $-$2.04451 & ~~~~~$JHK_{\rm s}$ & -- \\
b301 & 17:38:00.240 & $-$35:24:09.00 & 353.65635 & $-$2.04453 & $ZYJHK_{\rm s}$ & -- \\
b302 & 17:41:46.440 & $-$34:09:59.40 & 355.11565 & $-$2.04449 & $ZYJHK_{\rm s}$ & -- \\
b303 & 17:45:26.424 & $-$32:55:27.48 & 356.57487 & $-$2.04453 & $ZYJHK_{\rm s}$ & 1 \\
b304 & 17:49:00.600 & $-$31:40:34.32 & 358.03411 & $-$2.04446 & $ZYJHK_{\rm s}$ & 1 \\
b305 & 17:52:29.424 & $-$30:25:22.08 & 359.49334 & $-$2.04452 & $ZYJHK_{\rm s}$ & -- \\
b306 & 17:55:53.256 & $-$29:09:51.84 & 0.95261 & $-$2.04456 & $ZYJHK_{\rm s}$ & -- \\
b307 & 17:59:12.432 & $-$27:54:05.04 & 2.41186 & $-$2.04451 & $ZYJHK_{\rm s}$ & -- \\
b308 & 18:02:27.312 & $-$26:38:03.12 & 3.87111 & $-$2.04447 & $ZYJHK_{\rm s}$ & -- \\
b309 & 18:05:38.232 & $-$25:21:47.52 & 5.33034 & $-$2.04451 & $ZYJHK_{\rm s}$ & -- \\
b310 & 18:08:45.480 & $-$24:05:18.96 & 6.78962 & $-$2.04454 & $ZYJHK_{\rm s}$ & -- \\
b311 & 18:11:49.320 & $-$22:48:38.52 & 8.24891 & $-$2.04449 & $ZYJHK_{\rm s}$ & 1 \\
b312 & 18:14:50.040 & $-$21:31:47.64 & 9.70816 & $-$2.04447 & ~~~~~$JHK_{\rm s}$ & 1 \\
b313 & 17:25:32.232 & $-$37:14:49.92 & 350.73753 & $-$0.95236 & ~~~~~$JHK_{\rm s}$ & 1 \\
b314 & 17:29:34.800 & $-$36:02:05.64 & 352.19625 & $-$0.95230 & ~~~~~$JHK_{\rm s}$ & 1 \\
b315 & 17:33:30.168 & $-$34:48:52.92 & 353.65504 & $-$0.95232 & ~~~~~$JHK_{\rm s}$ & 1 \\
b316 & 17:37:18.768 & $-$33:35:14.28 & 355.11368 & $-$0.95231 & ~~~~~$JHK_{\rm s}$ & 1 \\
b317 & 17:41:01.104 & $-$32:21:10.80 & 356.57248 & $-$0.95229 & ~~~~~$JHK_{\rm s}$ & 1 \\
b318 & 17:44:37.584 & $-$31:06:45.00 & 358.03121 & $-$0.95230 & ~~~~~$JHK_{\rm s}$ & 1 \\
b319 & 17:48:08.616 & $-$29:51:58.32 & 359.48996 & $-$0.95233 & ~~~~~$JHK_{\rm s}$ & 1 \\
b320 & 17:51:34.560 & $-$28:36:52.56 & 0.94861 & $-$0.95235 & ~~~~~$JHK_{\rm s}$ & 1 \\
b321 & 17:54:55.800 & $-$27:21:28.44 & 2.40742 & $-$0.95234 & ~~~~~$JHK_{\rm s}$ & 1 \\
b322 & 17:58:12.648 & $-$26:05:48.12 & 3.86612 & $-$0.95234 & ~~~~~$JHK_{\rm s}$ & 1 \\
b323 & 18:01:25.416 & $-$24:49:52.68 & 5.32477 & $-$0.95232 & ~~~~~$JHK_{\rm s}$ & 1 \\
b324 & 18:04:34.440 & $-$23:33:42.84 & 6.78355 & $-$0.95232 & ~~~~~$JHK_{\rm s}$ & 1 \\
b325 & 18:07:39.984 & $-$22:17:20.40 & 8.24226 & $-$0.95238 & ~~~~~$JHK_{\rm s}$ & -- \\
b326 & 18:10:42.288 & $-$21:00:45.72 & 9.70101 & $-$0.95231 & ~~~~~$JHK_{\rm s}$ & -- \\
b327 & 17:21:01.656 & $-$36:37:48.00 & 350.73744 & 0.13984 & ~~~~~$JHK_{\rm s}$ & 1 \\
b328 & 17:25:06.528 & $-$35:25:37.20 & 352.19621 & 0.13989 & ~~~~~$JHK_{\rm s}$ & 1 \\
b329 & 17:29:04.152 & $-$34:12:56.52 & 353.65492 & 0.13987 & ~~~~~$JHK_{\rm s}$ & -- \\
b330 & 17:32:55.008 & $-$32:59:47.76 & 355.11368 & 0.13984 & ~~~~~$JHK_{\rm s}$ & -- \\
b331 & 17:36:39.504 & $-$31:46:13.08 & 356.57236 & 0.13988 & ~~~~~$JHK_{\rm s}$ & -- \\
b332 & 17:40:18.120 & $-$30:32:14.28 & 358.03110 & 0.13982 & ~~~~~$JHK_{\rm s}$ & -- \\
b333 & 17:43:51.192 & $-$29:17:52.80 & 359.48985 & 0.13988 & ~~~~~$JHK_{\rm s}$ & -- \\
b334 & 17:47:19.128 & $-$28:03:10.80 & 0.94855 & 0.13988 & ~~~~~$JHK_{\rm s}$ &   -- \\
b335 & 17:50:42.288 & $-$26:48:09.36 & 2.40731 & 0.13985 & ~~~~~$JHK_{\rm s}$ & -- \\
b336 & 17:54:00.984 & $-$25:32:49.92 & 3.86610 & 0.13985 & ~~~~~$JHK_{\rm s}$ & -- \\
b337 & 17:57:15.528 & $-$24:17:14.28 & 5.32478 & 0.13981 & ~~~~~$JHK_{\rm s}$ & -- \\
b338 & 18:00:26.208 & $-$23:01:23.16 & 6.78348 & 0.13983 & ~~~~~$JHK_{\rm s}$ & -- \\
b339 & 18:03:33.336 & $-$21:45:17.64 & 8.24226 & 0.13983 & ~~~~~$JHK_{\rm s}$ & 1 \\
b340 & 18:06:37.152 & $-$20:28:59.16 & 9.70099 & 0.13985 & $ZYJHK_{\rm s}$ & 1 \\
b341 & 17:16:35.472 & $-$36:00:07.56 & 350.73765 & 1.23203 & $ZYJHK_{\rm s}$ & 1 \\
b342 & 17:20:42.432 & $-$34:48:30.24 & 352.19686 & 1.23205 & ~~~~~$JHK_{\rm s}$ & 1 \\
b343 & 17:24:42.144 & $-$33:36:20.88 & 353.65613 & 1.23203 & ~~~~~$JHK_{\rm s}$ & 1 \\
b344 & 17:28:35.040 & $-$32:23:41.64 & 355.11542 & 1.23207 & ~~~~~$JHK_{\rm s}$ & 1 \\
b345 & 17:32:21.576 & $-$31:10:35.04 & 356.57468 & 1.23205 & ~~~~~$JHK_{\rm s}$ & 1 \\
b346 & 17:36:02.160 & $-$29:57:02.52 & 358.03399 & 1.23203 & ~~~~~$JHK_{\rm s}$ & 1 \\
b347 & 17:39:37.152 & $-$28:43:06.24 & 359.49322 & 1.23206 & ~~~~~$JHK_{\rm s}$ & 1 \\
b348 & 17:43:06.960 & $-$27:28:47.64 & 0.95251 & 1.23203 & ~~~~~$JHK_{\rm s}$ & 1 \\
b349 & 17:46:31.896 & $-$26:14:08.52 & 2.41172 & 1.23201 & ~~~~~$JHK_{\rm s}$ & 1 \\
b350 & 17:49:52.296 & $-$24:59:09.96 & 3.87096 & 1.23204 & ~~~~~$JHK_{\rm s}$ & 1 \\
b351 & 17:53:08.496 & $-$23:43:53.40 & 5.33027 & 1.23202 & ~~~~~$JHK_{\rm s}$ & -- \\
b352 & 17:56:20.760 & $-$22:28:20.28 & 6.78955 & 1.23201 & $ZYJHK_{\rm s}$ & -- \\
b353 & 17:59:29.376 & $-$21:12:31.68 & 8.24885 & 1.23202 & $ZYJHK_{\rm s}$ & -- \\
b354 & 18:02:34.608 & $-$19:56:29.04 & 9.70808 & 1.23199 & $ZYJHK_{\rm s}$ & -- \\
b355 & 17:12:13.584 & $-$35:21:49.68 & 350.73827 & 2.32427 & $ZYJHK_{\rm s}$ & 1 \\
b356 & 17:16:22.488 & $-$34:10:45.12 & 352.19857 & 2.32417 & $ZY$ & 1 \\
b357 & 17:20:24.096 & $-$32:59:06.36 & 353.65894 & 2.32423 & $ZY$ & -- \\
b358 & 17:24:18.888 & $-$31:46:56.64 & 355.11924 & 2.32422 & $ZY$ & -- \\
b361 & 17:35:26.472 & $-$28:07:39.36 & 359.50024 & 2.32421 & $ZY$ & -- \\
b362 & 17:38:58.008 & $-$26:53:43.80 & 0.96059 & 2.32418 & $ZYJHK_{\rm s}$ & -- \\
b363 & 17:42:24.624 & $-$25:39:25.92 & 2.42098 & 2.32420 & $ZYJHK_{\rm s}$ & -- \\
b364 & 17:45:46.632 & $-$24:24:47.88 & 3.88124 & 2.32419 & $ZYJHK_{\rm s}$ & -- \\
b365 & 17:49:04.368 & $-$23:09:50.40 & 5.34158 & 2.32416 & $ZYJHK_{\rm s}$ & -- \\
b366 & 17:52:18.096 & $-$21:54:34.92 & 6.80192 & 2.32418 & ~~~~~$JHK_{\rm s}$ & 1 \\
b367 & 17:55:28.104 & $-$20:39:02.88 & 8.26224 & 2.32419 & ~~~~~$JHK_{\rm s}$ & 1 \\
b368 & 17:58:34.656 & $-$19:23:15.00 & 9.72265 & 2.32423 & ~~~~~$JHK_{\rm s}$ & 1 \\
b369 & 17:07:55.872 & $-$34:42:57.24 & 350.73892 & 3.41637 & $ZYJHK_{\rm s}$ & 1 \\
b370 & 17:12:06.480 & $-$33:32:24.36 & 352.20083 & 3.41644 & $ZYJHK_{\rm s}$ & 1 \\
b371 & 17:16:09.864 & $-$32:21:16.20 & 353.66280 & 3.41638 & $ZYJHK_{\rm s}$ & -- \\
b372 & 17:20:06.384 & $-$31:09:35.28 & 355.12466 & 3.41637 & $ZYJHK_{\rm s}$ & -- \\
b373 & 17:23:56.496 & $-$29:57:23.04 & 356.58663 & 3.41642 & $ZYJHK_{\rm s}$ & -- \\
b374 & 17:27:40.584 & $-$28:44:42.36 & 358.04853 & 3.41640 & $ZYJHK_{\rm s}$ & -- \\
b375 & 17:31:19.032 & $-$27:31:34.68 & 359.51046 & 3.41635 & $ZYJHK_{\rm s}$ & -- \\
b376 & 17:34:52.152 & $-$26:18:01.44 & 0.97243 & 3.41644 & $ZYJHK_{\rm s}$ & -- \\
b377 & 17:38:20.328 & $-$25:04:05.16 & 2.43434 & 3.41638 & $ZYJHK_{\rm s}$ & -- \\
b378 & 17:41:43.848 & $-$23:49:46.56 & 3.89631 & 3.41638 & $ZYJHK_{\rm s}$ & -- \\
b379 & 17:45:03.024 & $-$22:35:07.44 & 5.35828 & 3.41637 & $ZYJHK_{\rm s}$ & 1 \\
b380 & 17:48:18.120 & $-$21:20:09.24 & 6.82020 & 3.41640 & $ZYJHK_{\rm s}$ & 1 \\
b381 & 17:51:29.424 & $-$20:04:53.40 & 8.28204 & 3.41640 & ~~~~~$JHK_{\rm s}$ & 1 \\
b382 & 17:54:37.224 & $-$18:49:20.64 & 9.74399 & 3.41636 & ~~~~~$JHK_{\rm s}$ & 1 \\
b383 & 17:03:42.216 & $-$34:03:30.60 & 350.73979 & 4.50856 & $ZYJHK_{\rm s}$ & 1 \\
b384 & 17:07:54.408 & $-$32:53:29.40 & 352.20380 & 4.50857 & ~~~~~$JHK_{\rm s}$ & 1 \\
b385 & 17:11:59.352 & $-$31:42:51.12 & 353.66783 & 4.50858 & ~~~~~$JHK_{\rm s}$ & 1 \\
b386 & 17:15:57.480 & $-$30:31:37.92 & 355.13195 & 4.50860 & ~~~~~$JHK_{\rm s}$ & 1 \\
b387 & 17:19:49.176 & $-$29:19:52.68 & 356.59597 & 4.50856 & ~~~~~$JHK_{\rm s}$ & 1 \\
b388 & 17:23:34.824 & $-$28:07:36.84 & 358.06004 & 4.50857 & ~~~~~$JHK_{\rm s}$ & 1 \\
b389 & 17:27:14.784 & $-$26:54:52.56 & 359.52411 & 4.50859 & ~~~~~$JHK_{\rm s}$ & 1 \\
b390 & 17:30:45.384 & $-$25:43:05.52 & 0.96034 & 4.50854 & ~~~~~$JHK_{\rm s}$ & 1 \\
b391 & 17:34:15.096 & $-$24:29:30.12 & 2.42444 & 4.50852 & ~~~~~$JHK_{\rm s}$ & 1 \\
b392 & 17:37:40.080 & $-$23:15:31.32 & 3.88850 & 4.50858 & ~~~~~$JHK_{\rm s}$ & 1 \\
b393 & 17:41:00.672 & $-$22:01:10.92 & 5.35253 & 4.50853 & ~~~~~$JHK_{\rm s}$ & -- \\
b394 & 17:44:17.136 & $-$20:46:29.64 & 6.81666 & 4.50854 & ~~~~~$JHK_{\rm s}$ & -- \\
b395 & 17:47:29.736 & $-$19:31:29.28 & 8.28074 & 4.50857 & $ZYJHK_{\rm s}$ & -- \\
b396 & 17:50:38.736 & $-$18:16:11.28 & 9.74476 & 4.50855 & $ZYJHK_{\rm s}$ & -- \\
\end{longtable}
}

\end{appendix}

\end{document}